%% file: main.tex
\documentclass[
 reprint,
 superscriptaddress,
 groupedaddress,
 citeautoscript,
 showkeys,
 preprintnumbers,
 numerical,
 amsmath,amssymb,
 aps,
 prb,
floatfix
]{revtex4-1}

\include{preamble}

\include{acronyms}

\begin{document}

\title{Strain and screening: Optical properties of a small-diameter carbon nanotube from first principles}

\author{Christian Wagner}
\email{christian.wagner@zfm.tu-chemnitz.de}
\affiliation{Technische Universit\"at Chemnitz, Center for Microtechnologies, Reichenhainer Stra\ss e 70, 09126 Chemnitz, Germany}
\affiliation{Department of Materials Science and Engineering, University of Illinois at Urbana-Champaign, Urbana, IL 61801, USA}
\affiliation{now at Helmholtz-Zentrum Dresden-Rossendorf, Bautzner Landstrasse 400, 01328 Dresden, Germany}

\author{J\"org Schuster}
\affiliation{Fraunhofer Institute for Electronic Nano Systems (ENAS), Technologiecampus 3, 09126 Chemnitz, Germany}

\author{Andr\'e Schleife}
\email{schleife@illinois.edu}
\affiliation{Department of Materials Science and Engineering, University of Illinois at Urbana-Champaign, Urbana, IL 61801, USA}
\affiliation{Frederick Seitz Materials Research Laboratory, University of Illinois at Urbana-Champaign, Urbana, IL 61801, USA}
\affiliation{National Center for Supercomputing Applications, University of Illinois at Urbana-Champaign, Urbana, IL 61801, USA}

\date{\today}

\begin{abstract}
Carbon nanotubes (CNTs) are a one-dimensional material system with intriguing physical properties that lead to emerging applications.
While CNTs are unusually strain resistant compared to bulk materials, their optical-absorption spectrum is highly strain dependent. 
It is an open question, as to what extent this is attributed to strain-dependent (i) electronic single-particle transitions, (ii) dielectric screening, or (iii) atomic geometries including CNT radii.
We use cutting-edge theoretical spectroscopy to explain strain-dependent electronic structure and optical properties of an (8,0) CNT.
Quasiparticle effects are taken into account using Hedin's $GW$ approximation and excitonic effects are described by solving a Bethe-Salpeter-equation for the optical polarization function.
This accurate first-principles approach allows us to identify an influence of strain on screening of the Coulomb electron-electron interaction and to quantify the impact on electronic structure and optical absorption of one-dimensional systems.
We interpret our thoroughly converged results using an existing scaling relation and extend the use of this relation to strained CNTs:
We show that it captures optical absorption with satisfactory accuracy, as long as screening, quasiparticle gap, and effective electron and hole masses of the strained CNT are known.
\end{abstract}


\maketitle

\section{Introduction}

\Acp{CNT} possess interesting material properties:
Their mechanical behavior is dominated by high stiffness and large rupture strain \cite{Lu_2005,Wu_2008, Zhang_2008,Park_2009,Pozrikidis_2009}, they are chemically very stable \cite{Hirsch_2002,Foerster_2015}, and show a sizable shift of electronic energy levels as a function of axial strain \cite{Yang_2000,Kleiner_2001,Minot_2003,Wagner_2012,Wagner_2016}.
This shift renders \emph{optical} transitions sensitive to strain, as has been observed experimentally \cite{Gopinath_2007,Maki_2007,Leeuw_2008,Huang_2008} and explained theoretically \cite{Spataru_2013}.
For this reason, \acp{CNT} are excellent candidates for electronic and optical strain sensing and optical strain characterization, which is a promising technique due to the practical ease of optical readout and the higher precision compared to alternative approaches such as indirect electronic characterization.
In particular, optical strain sensors with extremely high, mechanically tunable sensitivity can be built in combined \acs{CNT}/\ac{MOEMS} \cite{Maki_2007,Hierold_2008,Burg_2011,Helbling_2011}.

Unfortunately, there is no simple, quantitative picture of the explicit strain behavior of optical transitions, since their dependence on the single-particle band gap of the \acs{CNT} is not straightforward \cite{Spataru_2013}.
This can partly be attributed to strong many-body effects:
In low-dimensional systems such as quasi-\ac{1D} \acp{CNT}, there is less surrounding material than in bulk systems, leading to weak dielectric screening of the electron-electron and electron-hole interaction.
As a consequence, \ac{QP} shifts can be as large as 1.2 eV and excitonic effects can be equally strong \cite{Spataru_2004, Pedersen_2004, Zhao_2004, Deslippe_2009, Malic_2010, Spataru_2013}.
However, in order to achieve precise strain sensing based on \acp{CNT}, a thorough understanding of electronic and optical properties, as well as their strain dependence, needs to be developed.
Quantitative insight is essential for the development of \ac{MOEMS}, such as strain-tunable emitters based on \acp{CNT} or tunable optical sensors.

On a more fundamental level, \acp{CNT} are a well-suited test bed for obtaining deeper insight into the physics of the strain dependence of screening and, hence, the screened Coulomb electron-electron interaction $W$.
Understanding this is important for modern many-body perturbation theory, since in \ac{GW} and \ac{BSE} calculations, $W$ plays a crucial role for the renormalization of electronic QP energies and optical transition energies.
Large deformations are possible in \acp{CNT} before rupture, which allows exploring a much larger strain range than in bulk materials.

This understanding is also needed since difficulties often arise during the interpretation of experiments, e.g.\ for exciton binding energies:
While in (homogeneous) bulk material, a spatial average is a good approximation that describes dielectric screening using a dielectric constant $\epsilon$, this cannot \emph{a priori} be assumed for \acp{CNT}.
The spatially resolved dielectric function $\epsilon(\vec{r},\vec{r}')$ is needed because the material response, i.e.\ screening, is restricted to the actual electron density of the \acs{CNT} \cite{Deslippe_2009} and is, thus, strongly direction dependent.
In reciprocal space, spatial resolution corresponds to a dependence on $\vq$=$\vk-\vk'$, which means that $\epsilon(\vq$=$\vk-\vk')$ must be considered instead of a constant $\epsilon$.
Dynamical screening is captured by the frequency-dependent dielectric function $\epsilon(\vq,\omega)$, which is required when energy-dependent integrals occur.

Furthermore, as \ac{1D} materials, \acp{CNT} show negligible optical response perpendicular to the \ac{CNT} axis, i.e.\ along the $z$ direction.
Hence, screening $\epsilon(q_z$=$k-k')$  with $\vq$=$q_z \vec{e}_z$ and \ac{BZ} sampling are effectively \ac{1D}.
This needs to be taken into account when using analytical model functions to describe dielectric screening, since usually their $\vq$-dependence is fitted to \ac{3D} semiconductors with  a dielectric constant as low-$q$ limit \cite{Cappellini_1993}.
However, in \acp{CNT} the low-$q$ (large distance) limit is vacuum screening \cite{Jiang_2007, Deslippe_2009}.
Therefore, the $q$-dependence of $\epsilon$ and the emerging local-field effects must be calculated accordingly.
Nevertheless, in many studies only the dielectric constant $\epsilon$ is used as a screening model for the description of excitons in \acp{CNT}, since it is a much simpler quantity \cite{Pedersen_2003,Pedersen_2004,Perebeinos_2004}.
This neglect of local fields for the description of screening and the scaling of excitons in \acp{CNT} with respect to their radius, as proposed by Perebeinos \acs{etal} \cite{Perebeinos_2004}, is an approximation that requires careful revision.

In this work we use first-principles electronic-structure calculations to provide a deeper understanding of these questions.
We use \acs{DFT} \cite{Hohenberg_1964,Kohn_1965} to compute ground-state geometries and total energies of a small-diameter (8,0)-\acs{CNT} in equilibrium and under axial strain.
Hedin's $GW$ approximation \cite{Hedin_1965} is used to account for QP effects on electronic energy levels.
Using the $G_0W_0$ approximation, we derive strain-induced shifts of valence- and conduction-band energies and compare to results from a computationally cheaper hybrid exchange-correlation functional.
Finally, by solving the \ac{BSE} for the optical polarization function \cite{Onida_2002} we account for excitonic effects in optical-absorption spectra.
We study the influence of Coulomb truncation, a scheme used to mitigate finite-size effects in supercell calculations for low-dimensional systems, on resulting optical spectra of the \ac{CNT} under axial strain.

These detailed calculations of optical transitions allow us to disentangle the influence of strain on QP energies and on excitonic effects.
Using our data we explore whether the scaling relation by Perebeinos \acs{etal} \cite{Perebeinos_2004} for the exciton binding energy in different \acp{CNT} also holds for strain in a \acs{CNT}.
Finally, the relation between exciton binding energy, reduced effective mass, and dielectric constant is explored.
The resulting strain dependencies of exciton binding energies and optical transitions are essential ingredients for design and layout of \ac{MOEMS}.

The remainder of this work is structured as follows:
Section \ref{sec:theory} summarizes technical aspects of \ac{DFT}, $GW$, and \ac{BSE} calculations.
In Sec.\ \ref{sec:results} we use these techniques to discuss the strain-dependent electronic structure based on $G_0W_0$ and hybrid-functional calculations.
The solution of the \ac{BSE}
is shown and exciton binding energies are analyzed.
We then revisit the scaling relation of Ref.\ \onlinecite{Perebeinos_2004} and explore its applicability for the Coulomb-truncated case of a strained CNT.
Finally, Sec.\ \ref{sec:conc} summarizes and concludes our work.

\section{\label{sec:theory}Theoretical Approach and Computational Details}

\subsection{Ground-state properties}

We use \acs{DFT} \cite{Hohenberg_1964,Kohn_1965} to compute total energies and, via minimization of Hellman-Feynman forces, optimized ground-state geometries of a (8,0)-\acs{CNT} in equilibrium as well as under axial strain.
For these calculations the \ac{LDA} is used to describe \acl{XC} \cite{Perdew_1981} and the electron-ion interaction is described using norm-conserving pseudopotentials based on the parametrization by von Barth and Car \cite{Dal_Corso_1993}.
Wave functions are expanded into a plane-wave basis up to a cutoff energy of 550 eV (40 Ry).
To ensure accuracy, we also tested a plane-wave cutoff of 1100 eV (80 Ry), for which total energies are converged up to 9 meV/electron (36 meV/atom).
In both cases, the resulting \ac{DFT} as well as \ac{QP} gaps agree within 20 meV, which we include in our error bars for QP energy calculations (see below).
All \ac{DFT} calculations are carried out using the Quantum Espresso code \cite{Giannozzi_2009}.

We construct a simulation cell that contains the (8,0)-\ac{CNT}, oriented along the $z$ axis and surrounded by vacuum in the other two directions.
We choose a supercell size of 19.5\,$\times$\,19.5\,$\times$\,4.26 \AA$^3$, such that two periodic images of \acp{CNT} are separated by 13.2 \AA{}.
This is by far enough vacuum to obtain converged results and to suppress finite-size effects in \ac{DFT} calculations for the neutral \ac{CNT}.
The geometry optimization is performed using a 1\,$\times$\,1\,$\times$\,20 \ac{MP} \cite{Monkhorst_1976} ${\vk}$-point grid and all atoms are relaxed until the remaining forces are smaller than 0.01\,eV/\AA{}.
All our results for relaxed atomic geometries can be found in the supplemental material at [URL will be inserted by publisher].

We then compare to calculations within the Vienna \emph{Ab-initio} Simulation Package \cite{Kresse:1999,Kresse:1996}.
For these we use the generalized-gradient approximation by Perdew, Burke, and Ernzerhof \cite{Perdew_1996} and the projector-augmented wave method \cite{Bloechl:1994}.
The calculations are carried out using a plane-wave cutoff energy of 400\,eV and the same $\vk$-point grid discussed above.
Relaxed atomic geometries from both approaches differ only very slightly (GGA introduces about 0.2\,\% strain, see supplemental material at [URL will be inserted by publisher]), which is reassuring for the comparison of excited-state properties below.

\subsection{\textit{GW} calculations}

In order to describe QP effects on electronic single-particle energies, we use Hedin's $GW$ approximation for the electronic self energy \cite{Hedin_1965}.
We use the Yambo package \cite{Marini_2009} to compute QP energies within one step of perturbation theory, i.e.\ without updating $G$ or $W$, which is known as $G_0W_0$ approach.
The fully frequency-dependent dielectric response function $\epsilon(\vq,\omega)$, that enters $W$, is computed within \ac{RPA} using real-axis integration.
Local-field effects play an important role and are converged for a $\vG$-vector cutoff of 35\,eV ($\approx 2.4$\,Ry), which results in less than 0.5\,\% change of the dielectric function at several $\vq$-points with respect to the extrapolated value or less than 20 meV change in the $G_0W_0$ gap (see Fig.\ \ref{fig:conv_eps} in the supplemental material at [URL will be inserted by publisher]).
Converged calculations require a 1\,$\times$\,1\,$\times$\,40 \ac{MP} $\vk$-point grid (1\,$\times$\,1\,$\times$\,60 \ac{MP} $\vk$ points in case of Coulomb truncation, due to the sharper profile of $\epsilon(q_z)$, see Ref.\ \onlinecite{Spataru_2004}) and at least 256 bands, which is four times the number of occupied states.
Thus, the default parameter set is 256 bands and 60 ${\vk}$-points for \ac{GW} calculations, unless other values are explicitly given.
In addition, the singularity of the Coulomb integral in $\vk$ space has to be circumvented, which is achieved using the \ac{RIM} described by Marini \acs{etal}\cite{Marini_2009} for $GW$ calculations.

While the vacuum size in our supercell calculations is sufficiently large to achieve convergence in \ac{DFT}, a thorough unit cell convergence is not feasible for the screened Coulomb interaction $W$, due to its long-range character.
The slow decay of the Coulomb interaction with distance renders it impossible to eliminate artificial interactions between periodic images \cite{Rozzi_2006,Freysoldt_2008,Hueser_2013,Qiu_2016}.
This can be compensated, e.g.\ when calculating defect-formation energies, by subtracting the electrostatic contribution of all repeating cells \cite{Freysoldt_2008}.
For converged calculations of $W$, Coulomb truncation schemes were developed \cite{Spataru_2004,Rozzi_2006}, the Yambo implementation of which is used in this work and described in Ref.\ \onlinecite{Rozzi_2006}.
Using this scheme renders a lateral unit cell size of 19.5 \AA{} and a truncation cylinder for the Coulomb interaction (radius 9.75 \AA{}) sufficient.
All details on convergence tests for $\vk$ points, number of bands, cell size, and Coulomb truncation, including the non-trivial convergence studies with Yambo, are described in detail in the supplemental material at [URL will be inserted by publisher].

\subsection{Bethe-Salpeter calculations}

Excitonic effects are taken into account in the description of optical absorption by solving a \ac{BSE} for the optical polarization function \cite{Onida_2002}.
For \ac{BSE} calculations, the screened electron-hole interaction $W$ is computed using the static limit of the response function
and the same local-field effects as for $GW$ calculations \cite{Rohlfing_2000, Marini_2009}.
Ten valence and ten conduction bands are included for the solution of the \ac{BSE}.
Convergence with respect to $\vk$ points is achieved using a 2\,$\times$\,2\,$\times$\,80 \ac{MP} grid and the same lateral unit cell size of 19.5 \AA{} is found to be sufficient.
These convergence tests are discussed in detail in the supplemental material at [URL will be inserted by publisher]. 
The Yambo package (version 3.4.2) is used for all $GW$ and \ac{BSE} calculations.

In order to better understand the influence of dielectric screening, we compare Yambo results to \ac{BSE} calculations from a recent VASP-based implementation \cite{Roedl:2008,Fuchs_2008}.
These are carried out using the same parameters as for Yambo: 2\,$\times$\,2\,$\times$\,80 \ac{MP} $\vk$ points, ten valence and ten conduction bands, as well as the same simulation cell size.
Local-field effects are included up to 35\,eV $\vG$-vector cutoff, as discussed above for \ac{GW} calculations.
This allows calculating exciton binding energies that are converged to within about 1\,\% with respect to the dielectric $\vG$-vector cutoff. 
This error estimate stems from the extrapolation of the estimated error of the dielectric function (see Fig.\ \ref{fig:conv_eps} in the supplemental material at [URL will be inserted by publisher]).
An accurate extrapolation scheme is used to circumvent the Coulomb singularity \cite{Fuchs_2008}.
This BSE implementation currently does not support Coulomb truncation to remove the interaction between super cells, which is discussed in Sec.\ \ref{sec:optics_eps}.
In order to study the influence of the screened interaction $W$, we compare the results using a dielectric constant to an analytical model dielectric function \cite{Bechstedt:1992} for screening.
This comparison allows us to quantitatively discuss the interplay between Coulomb truncation, screening, and strain effects.

\section{\label{sec:results}Results and Discussion}

\subsection{Electronic structure of the unstrained (8,0)-CNT}

\begin{figure}
\includegraphics[width=0.99\columnwidth]{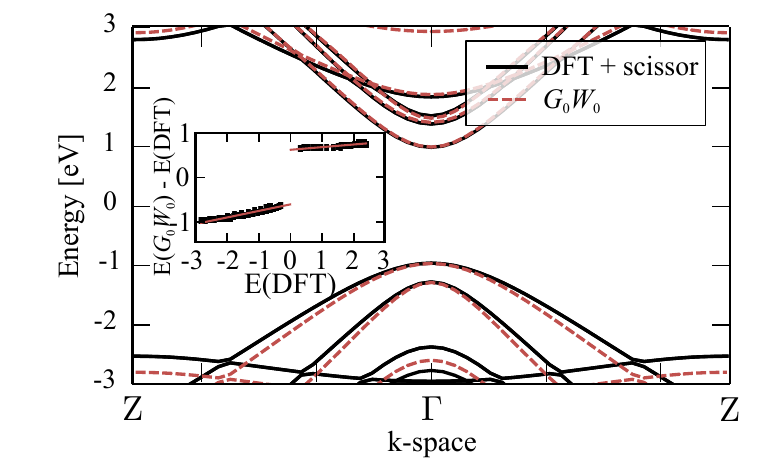}
\caption{\label{fig:dft_vs_gw}(Color online.) $G_0W_0$ band structure (red dashed) is compared to scissor-shifted \ac{DFT} results (black solid).
  Besides the scissor shift, the inset shows a linear dependence of \ac{QP} shifts on KS eigenvalues.
  The fit shows that the $GW$ valence band is stretched by about 1.15 and the $GW$ conduction band by about 1.06 with respect to corresponding KS bands.
  The Fermi level of the $GW$ band structure is chosen to be at zero energy.}
\end{figure}

In Fig.\ \ref{fig:dft_vs_gw} band structures computed using \ac{DFT} (scissor-shifted to 1.84\,eV) and $G_0W_0$ are compared.
Our $G_0W_0$ gap of 1.84$\pm$0.02\,eV (1.81$\pm$0.02\,eV when extrapolated to an infinite number of bands and ${\vk}$-points, see supplemental information [URL will be inserted by publisher]) agrees very well with early work by Spataru \acs{etal}, who reported 1.75 eV \cite{Spataru_2004}, and later work by Lanzillo \acs{etal}, who reported 1.81\,eV \cite{Lanzillo_2014}.
The small difference to Spataru \acs{etal} \cite{Spataru_2004} can be explained by slightly different computational parameters:
They used a plasmon-pole approximation to describe the $\omega$-dependence of the dielectric function and a slightly smaller 16\,\AA{} unit cell with Coulomb truncation beyond 7\,\AA{} cylinder radius.

The inset of Fig.\ \ref{fig:dft_vs_gw} shows that $G_0W_0$ QP shifts depend approximately linearly on DFT Kohn-Sham (KS) eigenvalues:
In addition to the scissor shift that opens up the gap, band stretching parameters $\beta$ describe the linear slope.
We find that \ac{VB} and \ac{CB} are stretched by $\beta\sub{vb}$=1.15 and $\beta\sub{cb}$=1.06.
This implies a small correction of effective masses,
$m^\textrm{GW}$=$\sqrt{\beta}\,m^\textrm{DFT}$, and needs to be taken into account when solving the \ac{BSE} with much finer $\vk{}$-point sampling.

In order to calculate effective masses of the $\pi$ bands, \footnote{Only the $\pi$ bands are optically active and their effective masses play an important role when comparing to results from \ac{BSE} calculations.} we use a hyperbolic fit that resembles the $G_0W_0$ bands as closely as possible \cite{Wagner_2016}.
The expression stems from the \ac{TB} zone folding approach together with the Dirac cone approximation \cite{Anantram_2006} for describing \ac{CNT} band structures.
The fit to \ac{DFT} data yields effective masses of $m\sub{cb}\up{DFT}$=0.422\,$m\sub{0}$ for the conduction and $m\sub{vb}\up{DFT}$=0.310\,$m\sub{0}$ for the valence band.
The effective masses of the respective $G_0W_0$ bands are $m\sub{cb}$=0.418\,$m\sub{0}$ and $m\sub{vb}$=0.278\,$m\sub{0}$, in quantitative agreement with band stretching.

\subsection{Hybrid functional for approximate QP energies}

The QP correction of the \ac{DFT} gap within the $G_0W_0$ approach is sizable:
The extrapolated shift is $1.21 \pm 0.02$ eV, compared to a \ac{DFT} gap of 0.60 eV.
This large shift is attributed to weak dielectric screening in the \ac{1D} \ac{CNT}, clearly indicating the need for using a QP correction scheme.
Unfortunately, the $G_0W_0$ approach is computationally expensive and becomes unaffordable, e.g.\ when a large number of \acp{CNT} or many different strained configurations are studied.
For these cases, an approximate description of QP corrections is beneficial and using a hybrid exchange-correlation functional, such as the one by Heyd, Scuseria, and Ernzerhof (HSE06) \cite{Heyd_2003,Heyd_2006,Krukau_2006} has proven successful.
It comes at much reduced computational cost, since no Coulomb truncation or expensive convergence with respect to empty states is needed.
For the (8,0)-\acs{CNT}, the reduction of cost is about a factor of $6$.

The HSE06 functional contains 25\,\% of \ac{HF} exact exchange and leads to a band gap of 1.06 eV for the (8,0)-\acs{CNT}.
Increasing the fraction of \ac{HF} exchange to 66\,\% reproduces the $G_0W_0$ band gap (see details in the supplemental material at [URL will be inserted by publisher]).
Such a large fraction of \ac{HF} exact exchange is not unusual for low-dimensional systems since screening is much weaker than in bulk materials, revealing almost bare electron-electron interaction.
Clearly, using a hybrid exchange-correlation functional \emph{without} adjusting the mixing parameter does not give correct band gaps for \acp{CNT}.
As an example, the work of Matsuda \acs{etal} publishes a band gap of about 1.28 eV for the (8,0)-CNT, using the B3LYP functional without adapting the mixing parameter\cite{Matsuda_2010}.
Next, we investigate whether the same fixed fraction of HF exchange results in sufficiently precise strain-dependent band gaps for the (8,0)-CNT, compared to $G_0W_0$ results.

\subsection{Electronic structure of the strained (8,0)-CNT}

\begin{figure}[!t]
\includegraphics[width=0.99\columnwidth]{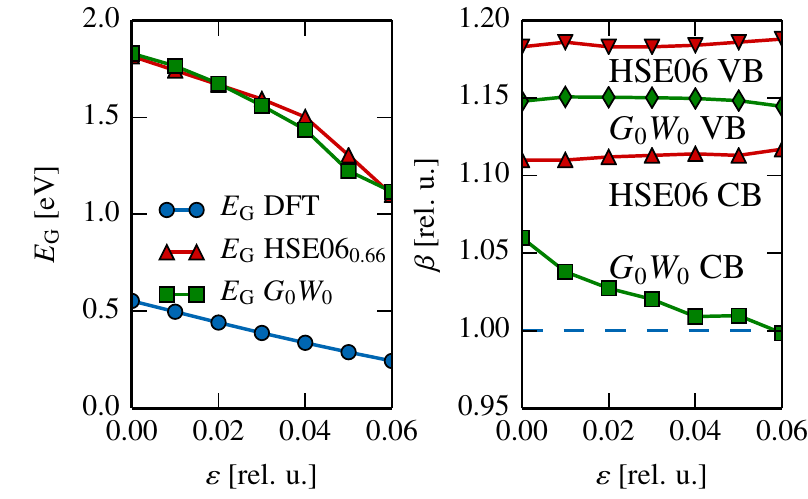}
\caption{\label{fig:gw_hse_bands_strain}(Color online.) Left: Band gaps of DFT, $G_0W_0$, and modified HSE06 calculations (66\,\% HF exact exchange) for the (8,0)-CNT under strain $\varepsilon$. 
Right: Band stretching parameters $\beta_\mathrm{CB,VB}$ of $G_0W_0$ and unmodified HSE06 calculations under strain (256 bands and 1\,$\times$\,1\,$\times$\,60 ${\vk}$ points). 
The blue dashed line indicates the DFT-LDA reference.}
\end{figure}

To investigate the strain dependence of the electronic structure, Fig.\ \ref{fig:gw_hse_bands_strain} shows fundamental gaps computed using DFT, modified HSE06, and $G_0W_0$, and band stretching parameters for several relative axial strains up to 6\,\%.
This illustrates the strong dependence of the fundamental gap on strain, which is significantly enhanced when QP effects are included, as seen from the different slopes of blue and green curves in the left panel of Fig.\ \ref{fig:gw_hse_bands_strain}.

This effect can be understood by invoking strain-dependent dielectric screening, in addition to strain-dependent shifts of KS eigenvalues computed in DFT (blue curve in Fig.\ \ref{fig:gw_hse_bands_strain}):
The smaller the band gap of the strained \acs{CNT}, the stronger the dielectric screening, and, thus, the weaker is the electron-electron repulsion.
Since QP shifts are small in a material with strong dielectric screening, the $G_0W_0$ gap of the \acs{CNT} with the largest axial strain (smallest gap) is closer to the DFT gap than for less strained \acp{CNT}.

Figure \ref{fig:gw_hse_bands_strain} also illustrates that the band gap computed using the modified HSE06 functional with 66\,\% exact exchange is very similar to the one computed using the $G_0W_0$ approach for all strains investigated here.
The remaining difference is less than 0.1 eV, showing that axial strains up to $\approx$\,6\,\% have no influence on the required amount of HF exchange.

We also note that while band stretching $\beta$
differs between \ac{CB} and \ac{VB}, it only slightly changes with strain:
$\beta\sub{cb}$ is reduced from 1.06 to 1.00 at 6\,\% strain and $\beta\sub{cb}$ remains at a constant value of 1.15.
The strain dependence of QP corrections modifies the strain-dependent effective mass of the \ac{CB} by less than 3\,\%.
The \ac{VB} is stretched by 15\,\% (7\,\% change of the effective mass), independent of the strain value.
Overall, this means that the ratio of \ac{DFT} and \ac{GW} corrected effective mass is close to 1.0 and, thus, barely strain dependent.
However, the absolute value of the effective mass (either from \ac{GW} or from \ac{DFT}) is strongly strain dependent as discussed in Sec.\ \ref{sec:exc_bind} (see Fig.\ \ref{fig:band_params_strain}).

\subsection{Optical properties of the strained (8,0)-CNT}

\begin{figure}[t]
\includegraphics[width=0.99\columnwidth]{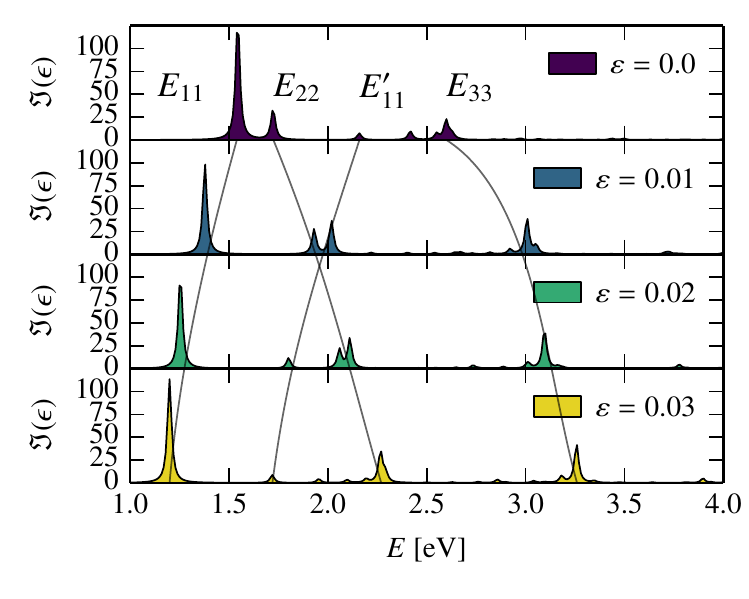}
\caption{\label{fig:result_bse_spectrum}(Color online.) Strain-dependent optical spectra of the (8,0)-CNT computed using the \ac{BSE} approach with Coulomb truncation.
  The $E_{11}$ and $E'_{11}$ transitions shift to lower energies, whereas $E_{22}$ and $E_{33}$ shift to higher energies.
  $E'_{11}$ denotes a higher-order exciton ($n$=2).
  Black lines are guides to the eye to highlight the shift of transitions.}
\end{figure}

\begin{figure}[t]
\includegraphics[width=0.99\columnwidth]{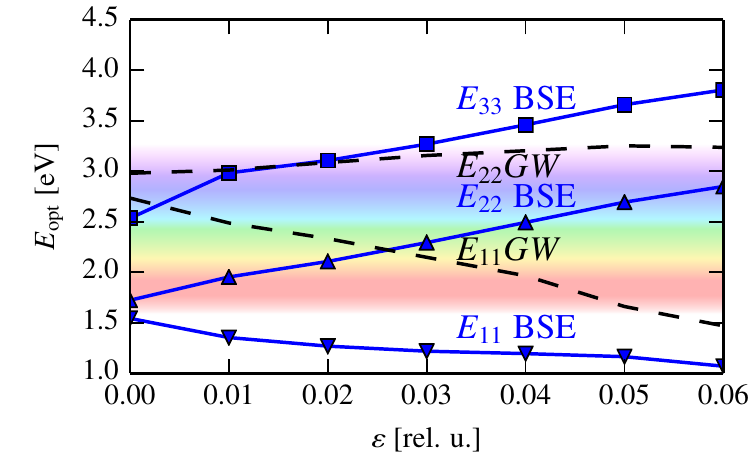}
\caption{\label{fig:bse_transitions_strain}(Color online.)
  Optical transition energies for the strained (8,0)-CNT with (BSE, blue solid) and without ($GW$, black dashed) excitonic effects, that shift strongly with applied axial strain, covering the visible spectral range.}
\end{figure}

We now discuss the strain dependence of the optical spectrum via strain-induced shifts of the transitions $E_{11}$, $E_{22}$, $E_{33}$, and $E'_{11}$ as depicted in Fig.\ \ref{fig:result_bse_spectrum}, where the index $nn$ indicates allowed transition from the $n^\mathrm{th}$ $\pi$-\ac{VB} to the $n^\mathrm{th}$ $\pi$-\ac{CB}\cite{Bachilo_2002}.
The unprimed transitions denote first-order excitons, whereas the primed transition $E'_{11}$ is a second-order exciton ($n$=2) that originates from the same electronic bands as $E_{11}$.
This assignment relies on the numerical diagonalization of the exciton Hamiltonian, whose eigenstates are superpositions of non-interacting KS states.
We analyzed these contributions for the different strained cases (see details in the supplemental material at [URL will be inserted by publisher]) and our assignment agrees with Spataru \acl{etal} for the unstrained case \cite{Spataru_2004}.
In the following, results with and without Coulomb truncation are discussed and the origin of the exciton binding energy is investigated.
The visualization of strain-dependent optical transitions in Fig.\ \ref{fig:bse_transitions_strain} shows that the first and second optical transition shift in opposite directions under strain.
This is consistent with the most simple \ac{TB} calculation of the CNT electronic bands with the zone-folding method applied to (strained) graphene \cite{Yang_2000}, which predicts a downshift of CNT bands with strain for odd transitions ($n$=$1,3,\dots$) and upshifts of even CNT bands ($n$=$2,4,\dots$).

In contrast to this \ac{TB} picture, we observe an upshift for the third optical transition that we attribute to $\sigma$-$\pi$-hybridization.
Since the curvature of the (8,0)-CNT is large, $\sigma$- and $\pi$-bands hybridize and the respective band energies are lowered.
This effect becomes stronger for higher bands and leads to reordering of the $n$=3 and $n$=4 states.
As a consequence, the third optical transition shifts in the direction opposite to what is predicted by the zone-folding model, which does not include an effect of a curved \ac{CNT} surface.

For the (8,0)-\acs{CNT}, the first optical transition $E_{11}$, which is often observed in photo- or electroluminescence, appears in the infrared and shifts towards lower energies.
For the unstrained \acs{CNT} we observe $E_{11}$ at 1.51$\pm$0.03 eV, which is nearly identical to 1.55 eV reported by Spataru \acs{etal} \cite{Spataru_2004}
We explain the small difference with the slightly different gaps, the use of \ac{RPA} instead of \ac{PPA}, and the slightly smaller unit cell.

Under strain, the $GW$+\ac{BSE} result for the $E_{11}$ transition shows a downshift to 1.02 eV at 6\,\% tensile strain.
Qualitatively, this trend follows the $GW$ results, but the exciton binding energy $E\sub{B}$, defined as difference between $GW$ (dashed black line in Fig.\ \ref{fig:bse_transitions_strain}) and $GW$+\ac{BSE} transition (blue line with markers in Fig.\ \ref{fig:bse_transitions_strain}), significantly reduces with strain.
We explain this via the strain-induced increase of the dielectric constant, i.e.\ screening (see Fig.\ \ref{fig:band_params_strain} and discussion in Sec.\ \ref{sec:exc_bind}), that leads to a reduction of the exciton-binding energy.

The $E_{22}$ and higher transitions are observed in optical absorption, \ac{PL} \cite{Bachilo_2002,Ohno_2006,Leeuw_2008}, Rayleigh scattering \cite{Wang_2006,Liu_2012}, and via photocurrents due to absorption \cite{Freitag_2003,Gabor_2012,Rauhut_2012}.
Figure \ref{fig:result_bse_spectrum} illustrates that $E_{22}$ and $E_{33}$ each consist of a series of peaks.
Their intensity-weighted average, depicted in Fig.\ \ref{fig:bse_transitions_strain}, shows that $E_{22}$ and $E_{33}$ shift approximately linearly in energy with strain by a large value of about 200 meV/\%.
For larger strained armchair \acp{CNT}, such as (11,0) and (17,0), that show a reduced $\sigma$-$\pi$-hybridization, this value is only about 150 meV/\% \cite{Spataru_2013}. 
Since these \acp{CNT} should possess about the same strain dependence within the \ac{TB} model with the zone folding scheme, it appears that the $\sigma$-$\pi$-hybridization itself is strain dependent.
This leads to an enhanced strain-sensitivity of electronic bands and corresponding optical transitions of the (8,0)-CNT.

\subsection{\label{sec:optics_eps}Exciton binding energies and long-range Coulomb interaction}

\begin{figure}[t]
\includegraphics[width=0.99\columnwidth]{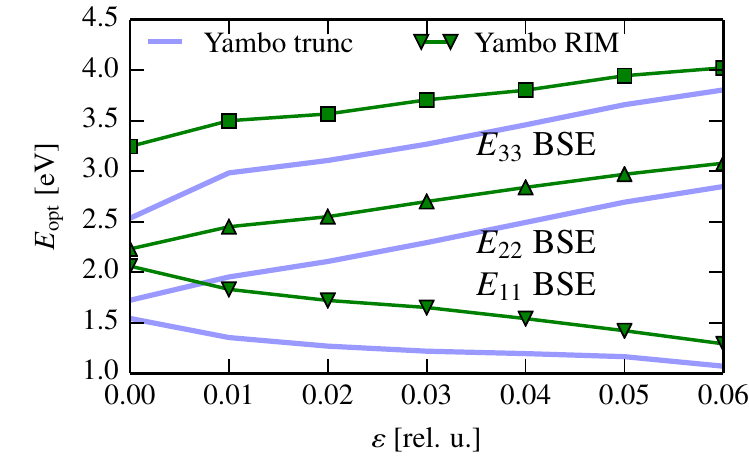}
\caption{\label{fig:bse_yambo_notrunc}(Color online.) The first three optical transitions computed with Yambo using \ac{RIM} and homogeneous screening (``Yambo RIM'') for the 19.5\,\AA{}\ unit cell. For comparison, the result with Coulomb truncation is included (``Yambo trunc'').}
\end{figure}

In order to understand the physics of screening in strained \acp{CNT}, we compare \ac{BSE} results with and without the Coulomb truncation scheme used for eliminating artificial Coulomb interactions of \acp{CNT} in adjacent super cells.
We use Yambo and the \ac{RIM} to solve the \ac{BSE} for the untruncated case and compare to the truncated case in Fig.\ \ref{fig:bse_yambo_notrunc}.
We also compare to the VASP-BSE implementation, which uses a different solver \cite{Fuchs_2008} and find that both codes agree almost perfectly, as documented in the supplemental material [URL will be inserted by the publisher].

Figure \ref{fig:bse_yambo_notrunc} shows that optical transitions appear at lower energies when the Coulomb interaction is truncated, which means that corresponding exciton binding energies are larger.
While in the untruncated case the electron-hole interaction is (artificially) affected by periodic images over long distances, in the truncated case, no periodic images are present and only the much smaller vacuum screening contributes.
Thus, the truncation affects the low-$q_z$ behavior of $\epsilon(q_z)$, which determines the screening of the electron-hole interaction in the long-range limit.
The reduction of screening for low $q_z$ due to Coulomb truncation explains the \emph{enhancement} of exciton binding energies.
Next, we establish detailed, quantitative insight into the scaling of exciton-binding energies with strain.

\subsection{\label{sec:exc_bind}Scaling of the exciton binding energy with strain}

Figure \ref{fig:bse_yambo_notrunc} also illustrates that the energies of optical transitions for untruncated and truncated cases depend on strain and  approach each other for large strain.
The reason is that the band gap is reduced with increasing strain, leading to increased screening that even becomes metallic for about 9\,\% strain.
In the metallic case, the truncation has almost no effect on the, then very large, screening \cite{Spataru_2004}.

This effect of strain-dependent screening on exciton binding energy and \ac{GW} gap does not just occur in \ac{1D} materials such as CNTs:
The exciton binding energy in bulk ZnO decreases from approximately 70 meV to 55 meV between $+2$\,\% and $-2$\,\% strain due to different screening, see Ref.\ \onlinecite{Schleife_2007}.
The effect is smaller in bulk, compared to low-dimensional systems, since screening is much stronger in 3D.
For various 2D materials with band gaps less than about 2 eV, where screening effects are almost as strong as in CNTs, Zhang \textit{et al.}\ showed that there is a simple, almost linear dependence of the exciton binding energy on the fundamental band gap \cite{Zhang_2017}.
They also showed that the absolute exciton binding energy is about 50\,\% of the band gap and reported that it changes as the band gap changes, e.g.\ due to strain.

\begin{figure}
\includegraphics[width=0.99\columnwidth]{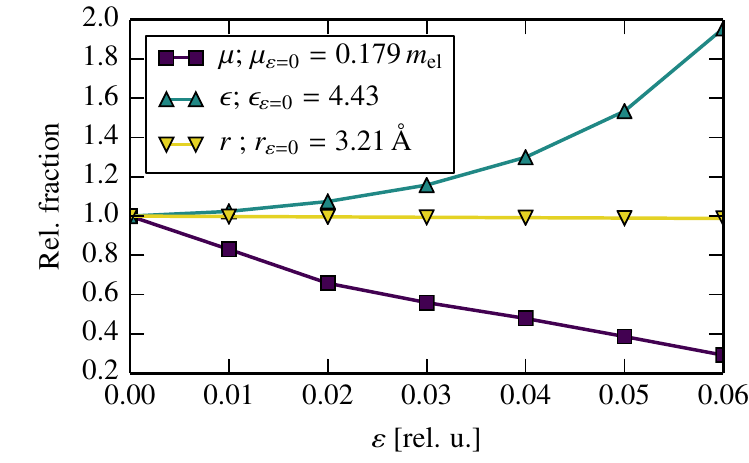}
\caption{\label{fig:band_params_strain}(Color online.) Strain-dependent reduced effective mass, dielectric constant (for the 19.5 \AA{} unit cell), and \ac{CNT} radius. These parameters enter the scaling relation for the exciton binding energy, Eq.\ \eqref{eq:bse_scaling}.}
\end{figure}

Next, we interpret the strain dependence of the exciton-binding energy via a scaling relation:
Perebeinos \acs{etal}\ derived this for CNTs, using a \ac{TB} Hamiltonian together with an Ohno potential in order to solve the \ac{BSE} \cite{Perebeinos_2004}.
By introducing a single parameter $\alpha$, they extended the well-known exciton scaling relation in homogeneous, isotropic materials \cite{Haug_2009}, $E\sub{B}\sim \mu\sub{eff}\epsilon^{-2}$, to
\begin{align}
\label{eq:bse_scaling}
  E\sub{B} \approx A\sub{B}\mu^{\alpha-1}\epsilon^{-\alpha}r\sub{CNT}^{\alpha-2},
\end{align}
where $A\sub{B}$ is the exciton-binding energy in a reference state, $r\sub{CNT}$ is the \ac{CNT} radius, $\mu$ the reduced mass of electron and hole, and $\epsilon$ is the dielectric constant.
Perebeinos \acs{etal}\ found a value of $\alpha$=1.40 for $\epsilon > 4$ for \acp{CNT}.
An independent confirmation of the parameter is given by Pedersen, who predicted a scaling of $E\sub{B}\sim r\sub{CNT}^{-0.6}$ using a variational approach for wave functions on a cylinder surface and homogeneous, background dielectric screening \cite{Pedersen_2003,Pedersen_2004}.
This result corresponds to the same value of $\alpha$=1.4 and $\sim r\sub{CNT}^{\alpha-2}$.
While the above relations were developed for a background dielectric screening, we now show that this screening (i.e.\ no Coulomb truncation) and local fields (i.e.\ with Coulomb truncation) are related.

To analyze the validity of this scaling relation for \acp{CNT} under strain, we depict our first-principles results for the three materials parameters that enter Eq.\ \eqref{eq:bse_scaling} in Fig.\ \ref{fig:band_params_strain}.
The dielectric constant is obtained from \ac{RPA} calculations using Yambo and the reduced effective mass results from our $G_0W_0$ data.
This figure shows that the \ac{CNT} radius depends only weakly on strain;
the Poisson ratio of about 0.2 leads to a shift in the exciton binding energy of about 0.7\,\% at 6\,\% tensile strain.
Conversely, the electronic structure is much more sensitive, leading to significant changes of effective masses and, via the fundamental gap, of the dielectric constant \cite{Yang_2000, Kleiner_2001, Wagner_2012, Wagner_2016}.
The two parameters $\mu$ and $\epsilon$, thus, determine the influence of strain on the exciton binding energy via Eq.\ \eqref{eq:bse_scaling}.

\begin{figure}
\includegraphics[width=0.99\columnwidth]{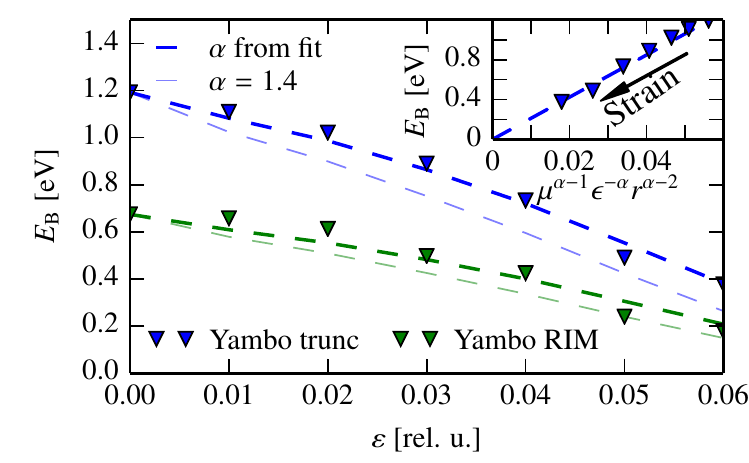}
\caption{\label{fig:bse_scaling_strain}(Color online.)
  Scaling of the strain-dependent exciton binding energy $E\sub{B}$ of the $E_{11}$ transition with (``Yambo trunc'') and without Coulomb truncation (``Yambo RIM''). 
Symbols represent BSE results and dashed lines represent the scaling relation, Eq.\ \eqref{eq:bse_scaling}, with different values of $\alpha$. The value of 1.40 given by Perebeinos \acs{etal}\cite{Perebeinos_2004} is compared to a fit to \ac{BSE} data.
Inset: Data for the Coulomb-truncated case as a function of the scaling parameter $\mu^{\alpha-1}\epsilon^{-\alpha}r^{\alpha-2}$.}
\end{figure}

In order to compare this to our \ac{BSE} results, we depict the strain-dependent exciton-binding energy of the $E_{11}$ transition in Fig.\ \ref{fig:bse_scaling_strain}.
This data is computed using the strain-dependent dielectric function $\epsilon(q_z)$ for screening of the electron-hole interaction and we compare results based on Coulomb truncation (see Fig.\ \ref{fig:eps_q_strain}) to those computed without the truncation scheme.
As expected, the resulting exciton binding energies differ in magnitude, since the underlying screening models deviate between truncated and non-truncated case, especially for low $q_z$ (see supplemental material [URL will be inserted by publisher]).

More importantly, Fig.\ \ref{fig:bse_scaling_strain} illustrates that the scaling relation, Eq.\ \eqref{eq:bse_scaling}, holds:
Fitting to results without Coulomb truncation yields a value of $\alpha\approx 1.29\pm 0.03$ and 
shows almost perfect agreement with our data, despite the fact that BSE calculations take local-field effects into account, whereas Eq.\ \eqref{eq:bse_scaling} was derived under the assumption of a constant, homogeneous dielectric screening.
Since there may be a significant influence from strong $\sigma$-$\pi$ hybridization due to CNT curvature, it is not surprising that the value of $\alpha$ slightly differs from 1.40 given by Perebeinos \cite{Perebeinos_2004}.
We note that the data in Fig.\ \ref{fig:bse_scaling_strain} was computed using Yambo; VASP data is shown in the supplemental material at [URL will be inserted by publisher].

Fitting to data with Coulomb truncation, yields a slightly different value of $\alpha \approx 1.21\pm 0.03$, since local-field effects with Coulomb truncation are not captured by the static, homogeneous screening entering the \ac{TB} models used by Pedersen\cite{Pedersen_2003,Pedersen_2004} or Perebeinos \cite{Perebeinos_2004}.
While this difference in $\alpha$ is, thus, not a surprise, it is remarkable that the scaling relation also holds in the Coulomb-truncated case, and we explore this in more detail in the next section.
We point out that for this fit, we used the dielectric constant from the untruncated case (see Fig.\ \ref{fig:band_params_strain}) to mimic background screening, since Coulomb truncation would imply $\epsilon=1.0$.
This is also addressed in the next section, where we introduce a geometry-dependent parameter $C_1$ to substitute $\epsilon$ in the scaling relation.
It characterizes the inhomogeneity and describes screening for confined carriers in the truncated geometry.

\subsection{Inhomogeneous dielectric screening and scaling relation}

\begin{figure}[t]
\includegraphics[width=0.99\columnwidth]{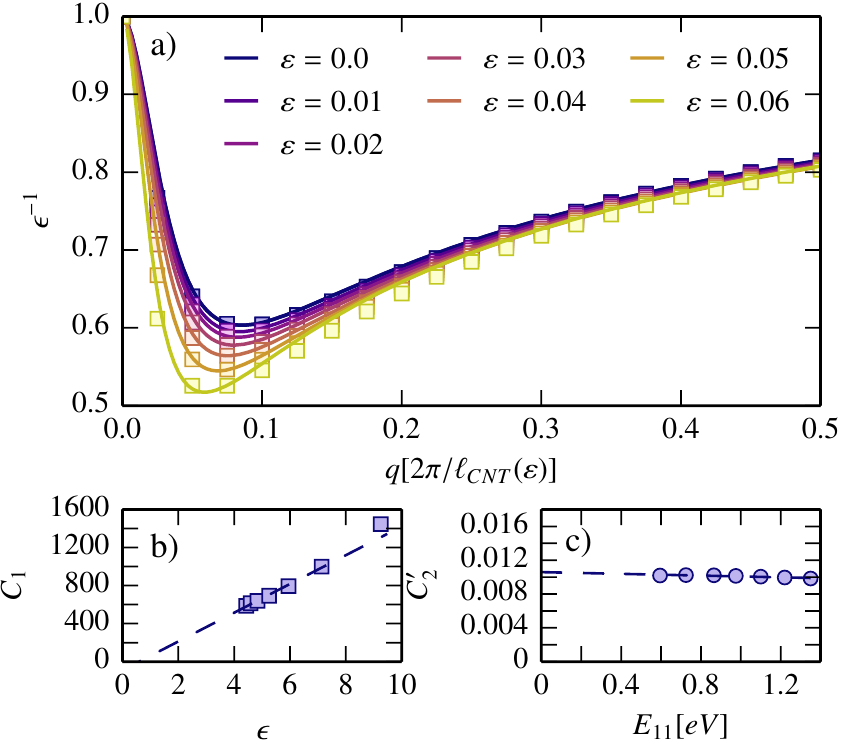}
\caption{\label{fig:eps_q_strain}(Color online.)
  In (a) the real part of the dielectric loss function $\epsilon^{-1}(q)$ is shown versus strain. 
  Squares represent first-principles data with Coulomb truncation and lines represent the model fit using Eq.\ \eqref{eq:eps_q_trunc}. 
The figures (b) and (c) show the fit coefficients $C_1$ and $C'_2$ as a function of $\epsilon$ and $E_{11}$, respectively.}
\end{figure}

For a single \ac{CNT}, as a localized, spatially inhomogeneous system, the wave-vector dependence of $\epsilon(q_z)$ is crucial when describing screening \cite{Spataru_2004,Deslippe_2009,Hueser_2013,Rasmussen_2016}.
In order to incorporate this into the scaling relation, we use the analytic expression for the dielectric function of an infinite \ac{1D} cylinder, derived by Deslippe \acs{etal} \cite{Deslippe_2009} using the Penn model \cite{Penn_1962}:
\begin{align}
\epsilon^{-1}\sub{1D}(q_z) &= 1 + \chi(q_z)\,v\sub{trunc}(q_z) \nonumber\\
 &\approx 1-C_2\frac{R}{E_{11}}\frac{C_1q_z^2}{1+C_1q_z^2} \sbr{2\,I_0(q_zR)\,K_0(q_zR)} \nonumber\\
 &= 1-C'_2R\frac{C_1q_z^2}{1+C_1q_z^2} \sbr{2\,I_0(qR)\,K_0(q_zR)}
 \label{eq:eps_q_trunc}
\end{align}
Here, $C_1$, $C_2$, and $C'_2=C_2/E_{11}$ are constants and $R$ is the \ac{CNT} radius. 
$I_0$ and $K_0$ are modified Bessel functions of the first and second kind, respectively.
We fit this expression to our first-principles data for $\epsilon(q_z)$ in Fig.\ \ref{fig:eps_q_strain} and observe very good agreement.
This means that the model of the \ac{1D} cylinder mimics screening in a CNT, once the influence of the supercell is removed via Coulomb truncation.

As shown in Fig.\ \ref{fig:eps_q_strain}, the resulting fit parameter $C_1$ linearly depends on the strain-dependent dielectric constant $\epsilon$; $C'_2$ is strain independent.
Therefore, $C_1$ carries the strain dependence of the screening function that was described by $\epsilon$ before Coulomb truncation was applied. 
The relation between $C_1$ and $\epsilon$ is almost linear, which explains why the scaling relation, Eq.\ \eqref{eq:bse_scaling}, also holds in the case of Coulomb truncation.
We can, therefore, rewrite Eq.\ \eqref{eq:bse_scaling} using $C_1$ instead of $\epsilon$:
\begin{align}
E\sub{B} = A\sub{B}r\sub{CNT}^{\alpha-2}\mu^{\alpha-1}C_1^{-\alpha}
\label{eq:bse_scaling_novel}
\end{align}
Hybrid DFT calculations can then yield effective masses and, after adjusting the fraction of exact exchange, strain-dependent corrected gaps.
In combination with the \ac{RPA}, these calculations also yield the inhomogeneous screening as a function of strain and, thus, the parameter $C_1$.
This shows that for an isolated, strained \ac{CNT}, the exciton-binding energy can be related to that of the unstrained state by means of a scaling relation, Eq.\ \eqref{eq:bse_scaling_novel}.

\section{\label{sec:conc}Conclusions}

We use first-principles electronic-structure calculations, based on the $GW$+\ac{BSE} approach, to compute strong, strain-related shifts of peaks $E\sub{nn}$ in the optical-absorption spectrum of an (8,0)-CNT, consistent with earlier literature.
We find that the exciton binding energy in strained \acp{CNT} is a function of the band gap and our work leads to the important conclusion that this arises directly from strain-dependent inhomogeneous dielectric screening.
This shows that deformation potentials of electronic eigenvalues \emph{and} exciton binding energies need to be considered explicitly, in order to predict strain-dependent optical spectra of \acp{CNT}.

While this implies that the effect of many-body physics on optical spectra in strained \acp{CNT} is crucial, we then show that a more simple scaling relation for the exciton binding energy is applicable also to strained CNTs.
This scaling relation allows us to extrapolate the shift of optical transitions from the unstrained state to the strained state, based on the strain-induced shift of electronic energy levels and the strain dependence of $\epsilon(q)$ and $\mu$.
We then showed that the modified HSE06 hybrid functional, with a fraction of 66\,\% exact exchange, mimics QP corrections for the unstrained \ac{CNT} quite well, allowing us to avoid expensive $GW$ calculations of strained \acp{CNT} to determine these parameters.

Finally, we provide detailed understanding of why the scaling relation works for strained CNTs, even though it relies on the dielectric constant as a parameter and neglects the influence of local-field effects.
To this end, we demonstrate that in low-dimensional materials, a wave-vector dependent screening function $\epsilon(q_z)$ must be used.
In addition, in first-principles excited-state calculations the Coulomb interaction must be truncated in order to obtain supercell convergence, which influences the long-range, low-$q_z$ part of the screening function.
We show that a suitable screening function $\epsilon(q_z)$ for \acp{CNT} can be obtained from a \ac{1D} Penn model of a charge on an infinitely long, hollow cylinder and connect the parameters of this model to our first-principles data, leading to an excellent fit.
We envision that this significantly advances the study of optical transitions in strained \acp{CNT} and enables broader applications of this interesting material system.

\begin{acknowledgments}
Part of this work was supported by the National Science Foundation under Grant No.\ DMR-1555153.
Support by the Deutsche Forschungsgemeinschaft DFG research unit 1713 and funding of the German academic exchange service (DAAD) is gratefully acknowledged.
We thank the Yambo developers, who gave important feedback concerning technical issues during some calculations.
\end{acknowledgments}

\end{document}


\title{Supplemental Material: Strain and screening: Optical properties of a small-diameter carbon nanotube from first principles}
\maketitle

\section{\label{app:eps_q_trunc}Dielectric function and local-field effects}

\begin{figure}[h]
\includegraphics[width=0.99\columnwidth]{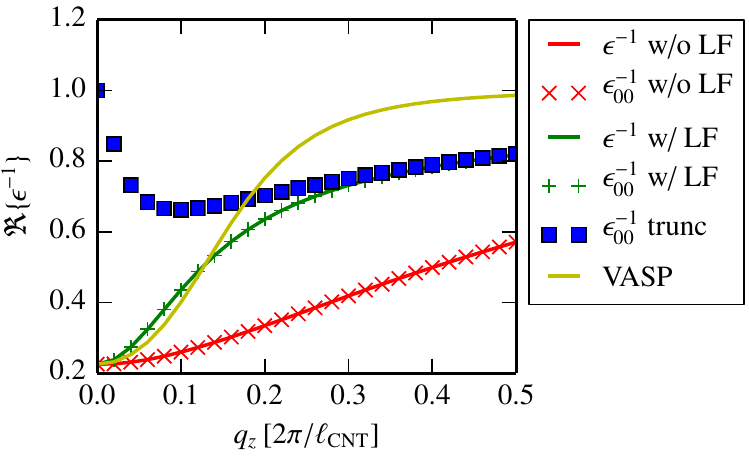}
\caption{\label{fig:eps_q_approx}(Color online.) $\epsilon^{-1}(q_z)$ for the (8,0)-CNT and the head element of the dielectric matrix $\epsilon^{-1}_{00}(q_z)$ without local-field effects, including local-field effects, and including Coulomb truncation. In the latter case, Yambo only provides $\epsilon^{-1}_{00}(q_z)$.}
\end{figure}

In order to compare our first-principles results for the screening function to the model function presented by Deslippe \cite{Deslippe_2009}, the dielectric loss function $\epsilon^{-1}(q_z)$ with Coulomb truncation is required.
Yambo, however, only exports the dielectric loss function in the periodic case, but provides the (large) dielectric loss matrix $\epsilon_{GG'}(q, \omega)$.
The dielectric loss function is calculated through:
\begin{align} \label{eq:LF}
\epsilon(q)=\lim\limits_{\vq'\rightarrow \vq}\frac{1}{\epsilon^{-1}_{00}(\vq', \omega)}
\end{align}
In order to correctly compute the head element of the dielectric matrix $\epsilon^{-1}_{00}$, Yambo evaluates $\epsilon_{GG'}=\delta_{GG'}+v_G(\vq)\chi_{GG'}(\vq, \omega)$ with the Coulomb potential $v_G(\vq)$ and $\chi_{GG'}=\chi^0_{GG'}(\vq,\omega)\sum_{G_1G_2}\chi^0_{GG_1}(\vq,\omega)v_{G_1}(\vq)$.
The susceptibility $\chi_{GG'}$ contains local-field corrections, whereas $\chi^0_{GG'}$ simply results from the independent-particle picture (without local-field effects).

In Fig.\ \ref{fig:eps_q_approx}, we validate the implemented post processing for the truncated $\epsilon^{-1}(q_z)$ by showing different levels of theory for $\epsilon^{-1}(q_z)$ of a CNT:
The case of $\epsilon^{-1}(q_z)$ without local-field effects (``$\epsilon^{-1}$ w/o LF'', red line) is compared to $\epsilon^{-1}(q_z)$ with local-field effects (``$\epsilon^{-1}$ w/ LF'', green line).
In addition, the head element of the dielectric loss matrix is depicted for these two cases: ``$\epsilon^{-1}_{00}$ w/ LF'' (green markers) and ``$\epsilon^{-1}_{00}$ w/o LF'' (red markers).''

\begin{figure*}[t]
\includegraphics[width=0.99\textwidth]{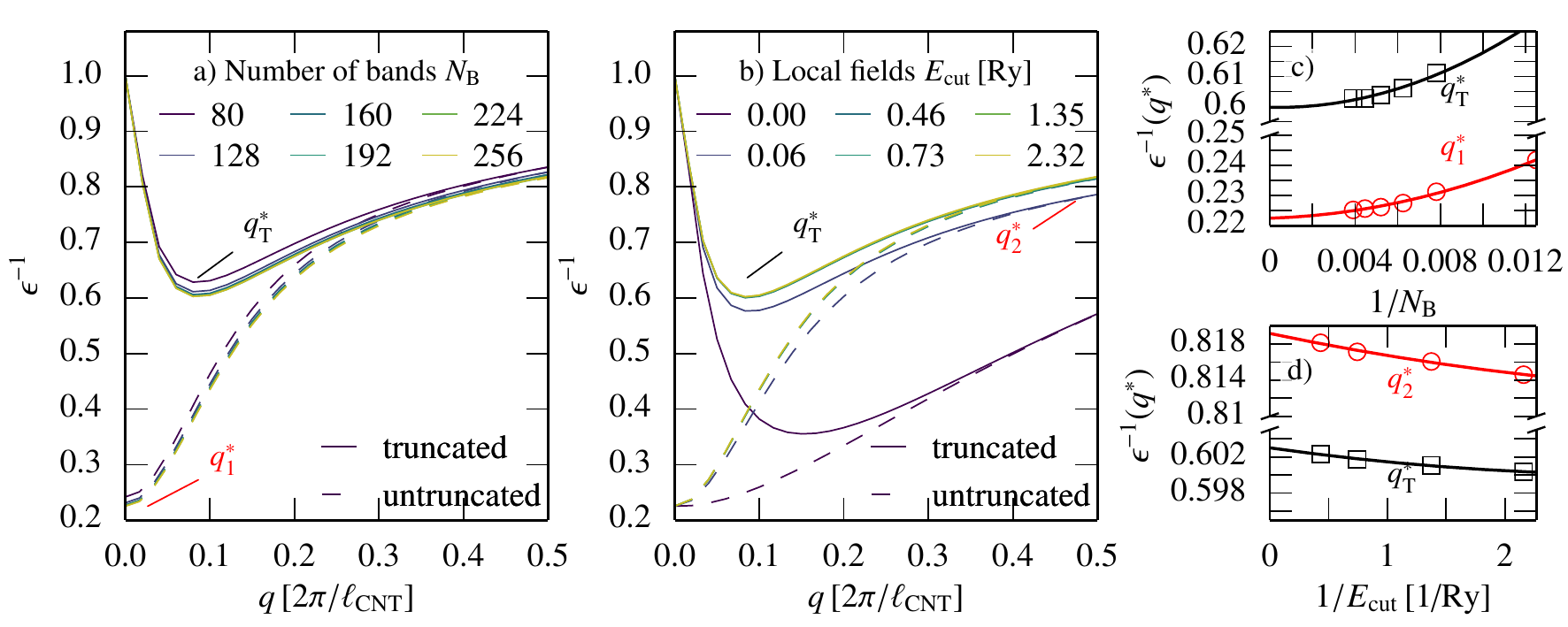}
\caption{Convergence of the dielectric loss function $\epsilon^{-1}(q)$ with the a) number of bands and b) local fields for a unit-cell size of 19.5\,\AA{}. Dashed lines do not include Coulomb truncation, whereas straight lines represent $\epsilon^{-1}(q)$ in the truncated case. Figures c) and d) quantify the change of $\epsilon^{-1}(q)$ at selected $q$-points: red labels are the standard case and black labels indicate the case with Coulomb truncation. Straight lines represent the extrapolation.} \label{fig:conv_eps}
\end{figure*}

\subsection{\label{app:eps}Convergence of the dielectric function}

We computed the dielectric function for different values of the local field $G$-vector cutoff and numbers of bands, both of which are convergence parameters (see Fig.\ \ref{fig:conv_eps}).
We first determined the value $q^\ast$, defined as the $\vq$ where $\epsilon(\vq)$ changes the most with respect to the convergence parameter. 
We then distinguish three cases:
For band convergence without Coulomb truncation, we use $q_1^\ast=0$ and, thus, compare dielectric constants.
For local-field convergence without Coulomb truncation, we study the high-$q$ limit, $q_2^\ast=q_\mathrm{max}$.
Since for the Coulomb-truncated case, $\epsilon(q=0)=1$ everywhere, we use the minimum of $\epsilon(q_\mathrm T^\ast)=\frac{1}{\epsilon_{00}^{-1}(q_\mathrm T^\ast)}$, where $\epsilon_{00}(q_\mathrm T^\ast)$ is the head element of the dielectric matrix at each $\vq$-point.
These three values for $q^\ast$ are illustrated in Fig.\ \ref{fig:conv_eps}.

In Fig.\ \ref{fig:conv_eps}, we also plot $\epsilon(\vq^\ast)$ as a function of the convergence parameters and extrapolate to infinity.
We then define the change of the dielectric function as the ratio of the numerically most expensive $\epsilon(\vq^\ast)^\mathrm{num}$ and the extrapolated value $\epsilon(\vq^\ast)^\mathrm{extrapol.}$.

\section{\label{app:gw}Convergence tests for the \textit{GW} calculations}

\subsection{Number of ${\vk}$-points}

\begin{figure}[h]
\includegraphics[width=0.99\columnwidth]{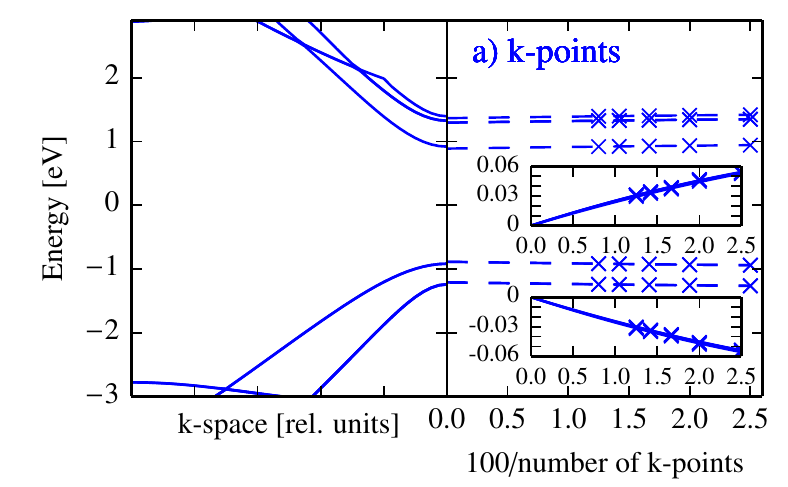}
\includegraphics[width=0.99\columnwidth]{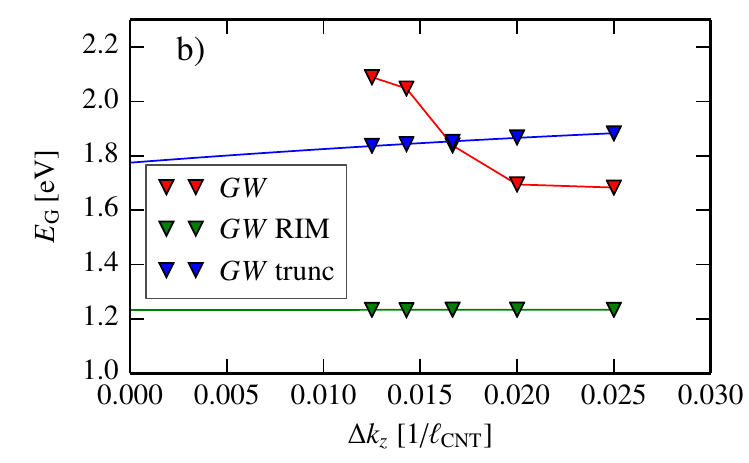}
\caption{\label{fig:gw_conv_k}(Color online.) 
Convergence of the energy bands with respect to the inverse number of $\vk$-points (256 bands, $10^5$ G-vectors).
a) The absolute ``$GW$ trunc'' eigenvalues as a function of $\Delta k_z$: The left panel show the band structure from $K$ to $\Gamma$ and the panel on the right side shows the band energies at $\Gamma$ as a function of the convergence parameter.
The insets show the highest \ac{VB} and lowest \ac{CB} as well as the extrapolated energy value for each band.
b) ``$GW$'' denotes standard $GW$ calculations, ``$GW$ RIM'' includes the \ac{RIM} for removing the Coulomb divergence, and ``$GW$ trunc'' uses the Coulomb truncation scheme. The 1x1x60 ${\vk}$-point set is considered converged (50 meV above the extrapolated value for an infinitely dense mesh, 1.79 eV). }
\end{figure}

The convergence of the $GW$ calculations with respect to the \ac{1D} $\vk$-mesh, presented in Fig.\ \ref{fig:gw_conv_k} is non-trivial.
We show three different setups: standard $GW$ calculations (``$GW$''), $GW$ including the \ac{RIM} for removing the divergence of the Coulomb integral (``$GW$ RIM''), and the case of Coulomb truncation (``$GW$ trunc'').
This shows that in the case of the ``standard'' $GW$ without \ac{RIM} or truncation, there is no convergence of the $GW$ gap.
It is already mentioned in the original publication by Marini \acs{etal}\ that the self-energy contribution suffers divergence due to the Coulomb singularity \cite{Marini_2009}.
Therefore, the \ac{RIM} is implemented and this method yields convergence for the band gap.
However, the value of the gap is far too small, which is due to insufficient convergence of the unit cell (see Fig.\ \ref{fig:gw_conv_cell}).
Activating Coulomb truncation of about half the cell size leads to convergence for the gap.
We note that higher bands converge in the same way as the band gap does (see Fig.\ \ref{fig:gw_conv_k}a and its inset).

However, after truncation, numerical parameters such as the number of $\vk$ points, the number of $G$ vectors in the exchange sum, and local-field effects have to be re-converged in order to get meaningful results.

\subsection{Convergence with respect to the number of $G$-vectors}

\begin{figure}[t]
\includegraphics[width=0.99\columnwidth]{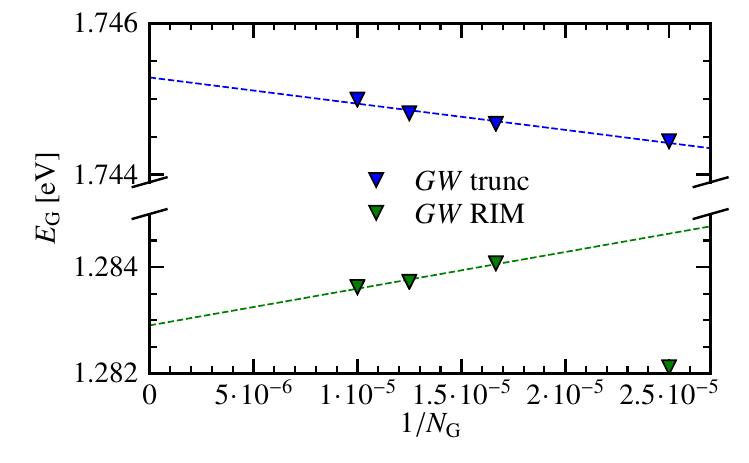}
\caption{\label{fig:gw_conv_G}(Color online.)
  The band gap as a function of the G-vector cutoff (number of G-vectors $N_G$).
``$GW$ RIM'' includes the \ac{RIM} for removing the Coulomb divergence and ``$GW$ trunc'' uses the Coulomb truncation scheme. 
$d_\mathrm{trunc}=15.5\,$\AA, $d_\mathrm{cell} = 19.5\,$\AA{}, (1$\times$1$\times$60 k-points, 256 bands)
}
\end{figure}

Figure \ref{fig:gw_conv_G} shows the band gap as a function of the inverse number of $G$ vectors for a given parameter set.
The value of the band gap is slightly smaller than in the other figures as the truncation radius was chosen to be $15.5\,$\AA{}.
From Fig.\ \ref{fig:gw_conv_G} the convergence is obvious and it can be seen that the $GW$ band gap is robust against minor changes of the number of $G$-vectors.
However, a minimum number of 60,000 $G$ vectors is required to obtain reliable values for the band gap.
If the value is lower, numerical noise and instabilities may occur during the $GW$ calculations as illustrated by the green data point at 2.5 in Fig.\ \ref{fig:gw_conv_G}, which corresponds to 40,000 $G$ vectors, and does not lie on the converging line.

\subsection{\label{app:gw_cell}Cell size convergence}

\begin{figure}[h]
\includegraphics[width=0.99\columnwidth]{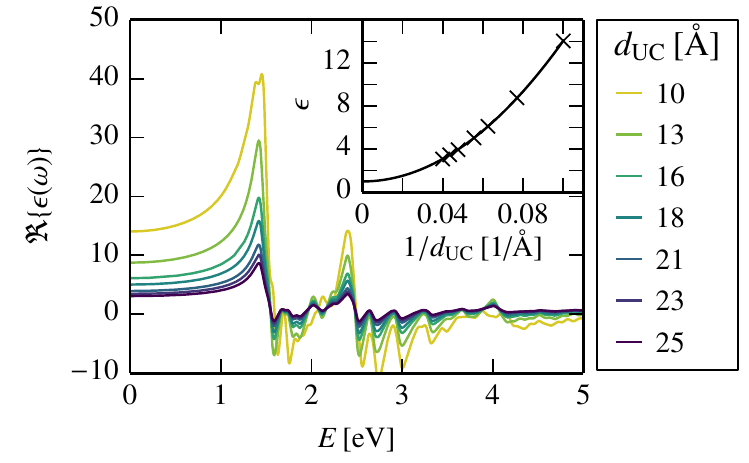}
\caption{\label{fig:cell_size_dft}
  (Color online.) Real part of the dielectric function as computed from \ac{RPA} based on the independent particle picture (\ac{DFT} results) for different unit-cell widths. 
The inset demonstrates that the dielectric constant $\epsilon$=$\epsilon_1(\omega$=$0)$ shows a $1/d\sub{cell}^2$ behavior (the fit function is a polynomial of order two) and converges towards the vacuum value of $\epsilon$=1 for infinite cell size.}
\end{figure}

Figure \ref{fig:cell_size_dft} shows the real part of the frequency-dependent dielectric function of the (8,0)-\ac{CNT} for different cell sizes in the \ac{IP} picture.
Clearly, the \emph{shape} of the dielectric function does not change for the cell sizes shown and we, thus, consider \ac{DFT} results converged for a unit cell width of 13 \AA.
The observed change in its magnitude simply originates from the normalization by unit-cell volume:
Increasing the simulation cell size by increasing the amount of vacuum modifies the $G$-space component $\epsilon_{GG}(q_z$=$0,\omega$=$0)\equiv\epsilon$ of the full dielectric matrix $\epsilon_{GG'}(q_z,\omega)$ in the \ac{IP} picture.
The inset of Fig.\ \ref{fig:cell_size_dft} shows that $\epsilon$ is reduced to vacuum screening for an infinite cell size and that local fields are responsible for the screening of the carriers.
This implies that the $q_z$-dependence of the screening must be included in $GW$ and BSE calculations.

\begin{figure}[h]
\includegraphics[width=0.99\columnwidth]{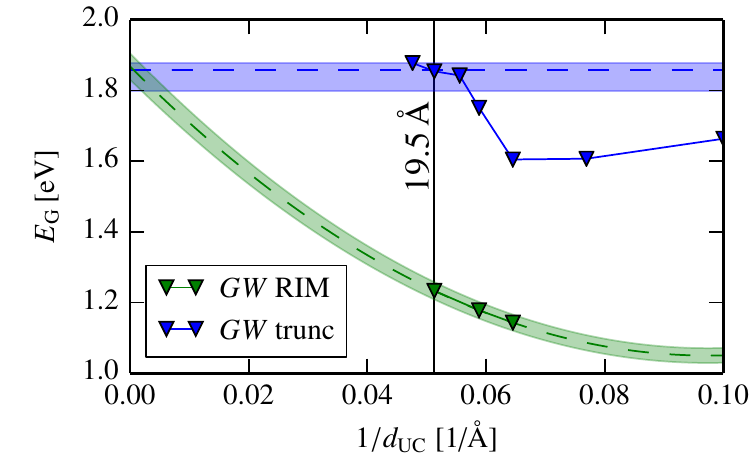}
\caption{\label{fig:gw_conv_cell}(Color online.) The convergence of the band gap with respect to the inverse supercell size in the $GW$ calculations: ``$GW$ RIM'' refers to  the application of the \ac{RIM} during the calculations, whereas ``$GW$ trunc'' refers to calculations with Coulomb truncation. The shaded area depicts the estimated error from the convergence test of the number of bands and the number of ${\vk}$-points. A cell size of about 19.5 \AA{} is considered converged.}
\end{figure}

This also means that in the case without Coulomb truncation, the cell average of the screening changes with the unit-cell size.
Figure \ref{fig:gw_conv_cell} depicts the $GW$ band gap with respect to the (inverse) cell size:
The band gap in $GW$ calculations with \ac{RIM}, i.e.\ without Coulomb truncation, considerably shifts for unit cells larger than 19 \AA.
Extrapolation to infinite cell size is possible, but the extrapolation error can be significant.

The validity of the extrapolation of the unit-cell size with \ac{RIM} can only be justified by the comparison to the Coulomb truncated case, since the application of Coulomb truncation \cite{Rozzi_2006} speeds up convergence tremendously (see Fig.\ \ref{fig:gw_conv_cell}).
This illustrates that the gap with Coulomb truncation has about the same value as the gap using \ac{RIM}, extrapolated to an infinite cell. 
The truncation range chosen here corresponds to the situation when about 99\,\% of charge density lies within the truncation cylinder, which must be (less than) half the unit-cell size.
If the truncation radius is significantly smaller, some parts of the electron density do not contribute to the screening.
The resulting gap in these cases is meaningless, as can be seen from the scattering of the band gap for smaller cell sizes.

\subsection{Interrelation of the convergence parameters}

The unit-cell size $d_\mathrm{UC}$, number of $G$-vectors $N_\mathrm{G}$, and number of bands $N_\mathrm{B}$ are interrelated convergence parameters and in the following we study how this impacts convergence of our band-gap results.
Extrapolation to infinite values of these convergence parameters is simplified by plotting inverse quantities and extrapolating to zero.
For this reason we investigate the change of the band gap $E_\mathrm{G}$ with respect to unit-cell size, using inverse cell size $d_\mathrm{UC}^{-1}$, inverse number of $G$-vectors $N_\mathrm{G}^{-1}$, and inverse number of bands $N_\mathrm{B}^{-1}$, and the expression
\begin{align}
  \label{eq:conv}
  \nonumber
  \Delta E_\mathrm{G}\approx&\frac{\dd E_\mathrm{G}}{\dd d_\mathrm{UC}^{-1}}\Delta d_\mathrm{UC}^{-1}= \parfrac{E_\mathrm{G}}{d_\mathrm{UC}^{-1}}\Delta d_\mathrm{UC}^{-1}\\ \nonumber
&+ \left(\parfrac{E_\mathrm{G}}{N_\mathrm{G}^{-1}}\parfrac{N_\mathrm{G}^{-1}}{d_\mathrm{UC}^{-1}} + \parfrac{E_\mathrm{G}}{N_\mathrm{B}^{-1}}\parfrac{N_\mathrm{B}^{-1}}{d_\mathrm{UC}^{-1}}\right)\Delta d_\mathrm{UC}^{-1}\\
\equiv&\parfrac{E_\mathrm{G}}{d_\mathrm{UC}^{-1}}\Delta d_\mathrm{UC}^{-1}+\Delta_\mathrm{inter}\Delta d_\mathrm{UC}^{-1}  .
\end{align}
To analyze the interrelation, we focus on $\Delta_\mathrm{inter}$ and evaluate explicit numerical values for the different terms from our convergence studies:
Figure \ref{fig:gw_conv_G} yields a slope of $\partial E_\mathrm{G}/\partial N_\mathrm{G}^{-1}\approx -0.4$\,meV$/10^{-5}$, i.e.\ the energy difference between $N_G^{-1}=2.0\times 10^{-5}$ and $N_G^{-1}=1.0\times 10^{-5}$.
From Fig.\ \ref{fig:gw_conv_bands}, we find an average slope of $\partial E_\mathrm{G}/\partial N_\mathrm{B}^{-1}\approx -10\,$meV$/0.004$, i.e.\ the energy difference between $N_\mathrm{B}^{-1}=1/128\approx 0.008$ and $N_\mathrm{B}^{-1}=1/256\approx 0.004$.

We then determine $\partial N_\mathrm{G}^{-1}/\partial d_\mathrm{UC}^{-1}$:
The number of $G$-vectors per volume is a measure for the resolution of the electron density.
Since two dimensions are simultaneously increased when the unit-cell size is increased, the number of required $G$-vectors is a quadratic function of the unit-cell size,
$N_\mathrm{G} = \alpha d_\mathrm{UC}^2$, which means that $\parfrac{N_\mathrm{G}^{-1}}{d_\mathrm{UC}^{-1}}=\frac{2}{\alpha} d_\mathrm{UC}^{-1}$.

\begin{figure}[b]
\includegraphics[width=0.49\columnwidth]{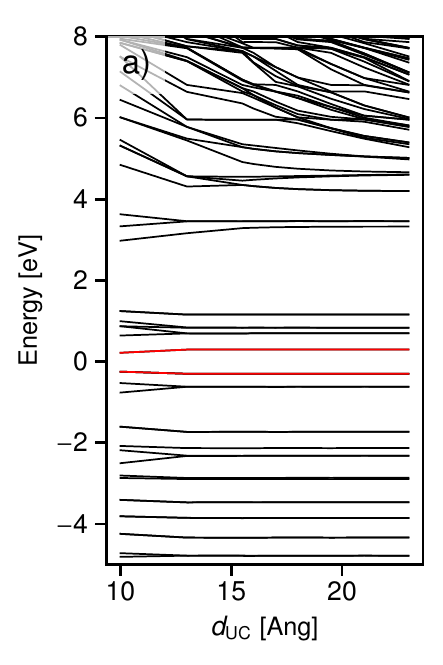}
\includegraphics[width=0.49\columnwidth]{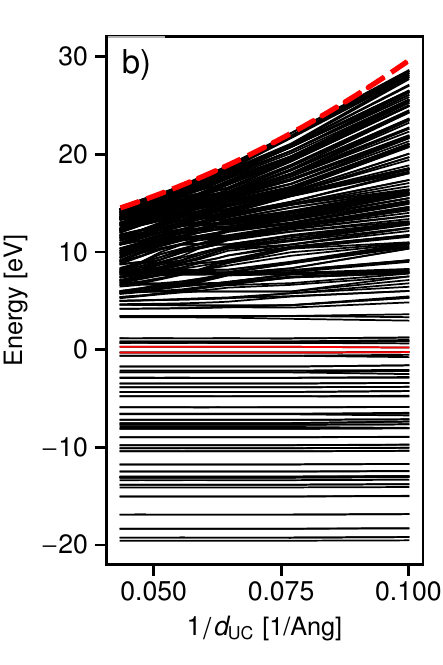}
\caption{\label{fig:dft_conv_cell}(Color online.)
  (a) DFT energy levels at the $\Gamma$ point are shown versus unit-cell size for a narrow energy window.
  (b) DFT energy levels at $\Gamma$ as a function of the inverse unit-cell size for 256 bands.
  Red solid lines depict valence-band maximum and conduction-band minimum.
  The red dashed line depicts the scaling of the highest band with the inverse uni-cell size.
}
\end{figure}

To obtain $\partial N_\mathrm{B}^{-1}/\partial d_\mathrm{UC}^{-1}$ we need to evaluate how many bands are needed, for a given unit-cell size, to include states up to a certain energy $E_\mathrm{max}$.
Figure \ref{fig:dft_conv_cell}(b) illustrates that this number depends on the unit-cell size, since the energy of states far above the Fermi level depends on the unit-cell size.
We verified that occupied states, which arise from CNT states, do not shift as a function of the cell size, unless too small cells are used, see Fig.\ \ref{fig:dft_conv_cell}(a).
For the highest energy level in a 2D box of length $d_\mathrm{UC}$ we know $E_\mathrm{max}=\beta N_\mathrm{B}^2/d_\mathrm{UC}^2$, i.e.\ $\partial N_\mathrm{B}^{-1}/ \partial d_\mathrm{UC}^{-1}=\sqrt{\beta / E_\mathrm{max}}$.
Figure \ref{fig:dft_conv_cell}(b) confirms this behavior.

All these terms together then lead to the following expression for $\Delta_\mathrm{inter}$:
\begin{align}
\Delta_\mathrm{inter} &=
-\frac{0.4\,\mathrm{meV}}{10^{-5}}\cdot{\frac{2}{\alpha} d_\mathrm{UC}^{-1}} - \frac{10\,\mathrm{meV}}{0.004}\cdot {\sqrt{\frac{\beta}{E_\mathrm{max}}}} \nonumber
\end{align}
We then substitute the values back in that we determined to be converged, i.e., $\alpha = N_{G,\mathrm {conv}}/d_\mathrm{UC,conv}^2$ and $\beta/E_\mathrm{max} = d_\mathrm{UC,conv}^2/N_\mathrm{B,conv}^2$, leading to
\begin{align}
\Delta_\mathrm{inter} &= -\frac{0.4\,\mathrm{meV}}{10^{-5}}\cdot{
 \frac{2\cdot N_{G,\mathrm{conv}}^{-1}\,d_\mathrm{UC}^{-1}}{d_\mathrm{UC,conv}^{-2}} } \nonumber \\ &-\frac{10\,\mathrm{meV}}{0.004}\cdot \frac{N_\mathrm{B,conv}^{-1}}{d_\mathrm{UC,conv}^{-1}}.
\end{align}
Here, $N_\mathrm{B,conv}=256$, $N_{G,\mathrm {conv}}=10^5$, and the reference cell size is 19.5\,\AA.
As we interpolate to the infinite unit cell, $\Delta d_\mathrm{UC}^{-1} = d_\mathrm{UC,conv}^{-1}$ and the explicit value of $d_\mathrm{UC,conv}$ does not appear.
Finally, we obtain:
\begin{align}
 \Rightarrow \Delta_\mathrm{inter} &= - 2\cdot 0.4\,\mathrm{meV} - 10\,\mathrm{meV} = -10.8\,\mathrm{meV}.
\end{align}

This change is rather small and within the theoretical error bar.
Thus, once all parameters are converged properly, the interrelations between the convergence parameters are only a small additional correction.

\subsection{Number of empty states}

\begin{figure}[t]
\includegraphics[width=0.99\columnwidth]{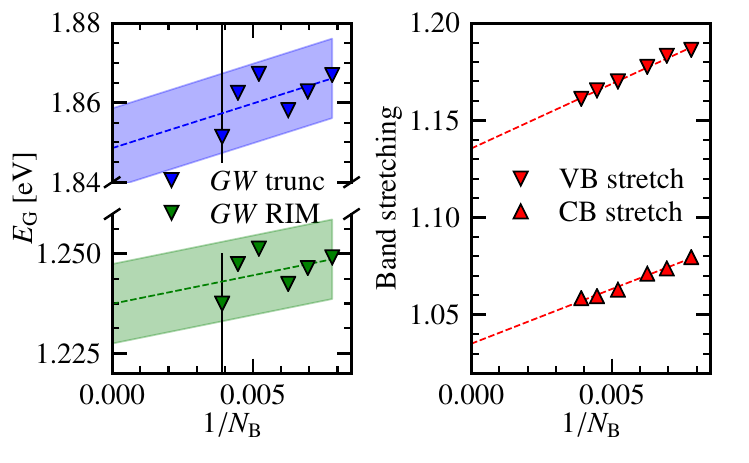}
\caption{\label{fig:gw_conv_bands}(Color online.) The dependence of the $G_0W_0$ band gap on the inverse number of electronic bands $N_\mathrm{B}$ is shown. 
In this work we use $N_\mathrm{B,conv}=256$ bands, which corresponds to less than 20 meV near the extrapolated value for an infinite number of bands. 
The band stretching parameters change insignificantly for the extrapolated values.}
\end{figure}

Since $GW$ results can sensitively depend on the number of empty states, we vary the number of conduction bands and extrapolate the results to an infinite number of bands, see Fig.\ \ref{fig:gw_conv_bands}.
The extrapolated gap and the gap for 256 bands differ by less than 20\,meV, which is why we use 256 bands throughout this work.

\subsection{Fit of the HSE functional to $GW$ calculations}

\begin{figure}[h]
\includegraphics[width=0.99\columnwidth]{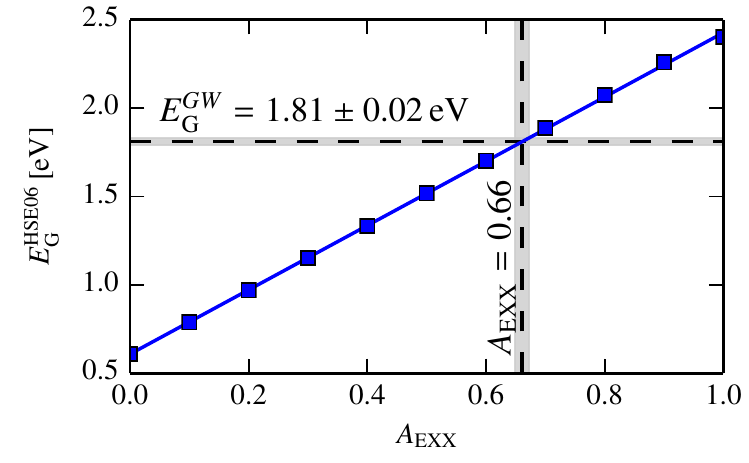}
\caption{\label{fig:HSE_fit_gw}(Color online.) HSE band gap of the (8,0)-CNT using different fractions of exact exchange ($A\sub{EXX}$). 
For the (8,0)-CNT, a value of $A\sub{EXX}=0.66\pm 0.01$ is required to match the (extrapolated) $GW$ gap of $1.81\pm0.02$\,eV. 
The default for HSE06 is 0.25.}
\end{figure}

For the calculation of QP effects, the HSE06 functional \cite{Heyd_2003,Heyd_2006,Krukau_2006} is applied using the implementation in VASP.
The fraction of exact exchange is altered in a range between 0 and 1.
It becomes apparent that the fraction of exact exchange to be added is rather large in order to reproduce the $GW$ band gap.
As mentioned in the main text, this is due to the incomplete screening in a \ac{1D} material, compared to bulk.

\section{\label{sec:bseconv}Convergence of the BSE calculations}

\subsection{\vk{}-point sampling}

We use different \vk{}-point samplings in order to converge the optical spectra:

Since VASP-BSE uses a singularity correction (see Ref.\ \onlinecite{Fuchs_2008}) that requires finite volume elements in 3D $\vk$ space, 2\,$\times$\,2\,$\times$\,$N$ ($N$ is the number of $\vk_z$-points) meshes are required to compare Yambo-BSE results with VASP-BSE under exactly the same conditions.
In Yambo, the use of a 1\,$\times$\,1\,$\times$\,$N$ mesh is possible, but in the following, we performed all tests with a 2\,$\times$\,2\,$\times$\,$N$ mesh.

\begin{figure}[h]
\includegraphics[width=0.99\columnwidth]{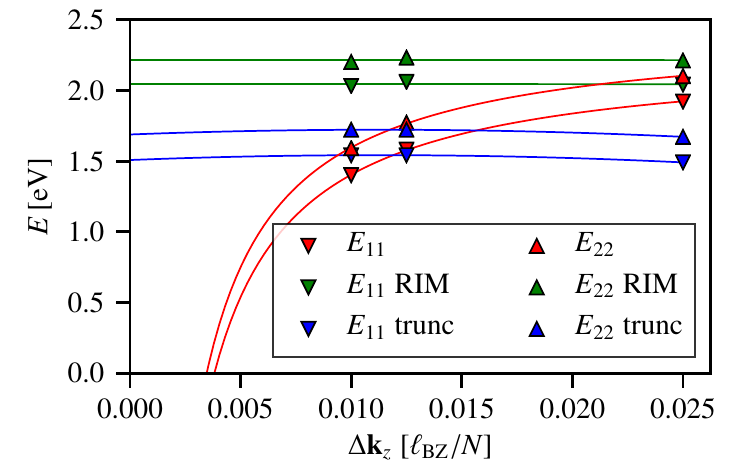}
\caption{\label{fig:bse_conv_dk}(Color online.) Convergence of the first optical transitions ($E_{11}$ and $E_{22}$) for different ${\vk}$-point samplings: In the case of the \ac{BSE} without \ac{RIM} or truncation (labeled $E_{11}$, $E_{22}$), the peaks do not converge.
In the truncated case (``cut'') and in the case with \acl{RIM} (``RIM''), there is convergence.}
\end{figure}

Solving the \ac{BSE} leads to convergence problems when increasing the mesh density, which is equivalent to decreasing $\Delta \vk_z=\ell_\mathrm{BZ}^z/N$.
Fig.\ \ref{fig:bse_conv_dk} visualizes that an increased density of \vk{}$_z$-points (decreasing $\Delta \vk_z$) leads to a continuous redshift of the first optical transitions $E_{11}$ and $E_{22}$ if neither Coulomb truncation nor singularity correction is applied.
The exchange part of the self energy diverges with increasing \vk{}$_z$-point density, which leads to this result \cite{Marini_2009}.

The \ac{RIM} that is implemented in the Yambo code resolves this problem and circumvents the divergence.
Convergence with respect to the $\vk$-point sampling can then be achieved (``$E_{11}$ RIM'' and ``$E_{22}$ RIM'' in Fig.\ \ref{fig:bse_conv_dk}).
Further, the spectrum converges in the case of the Coulomb truncation (``$E_{11}$ trunc'', ``$E_{22}$ trunc'').
This is due to the lifting of the singularity in the Coulomb interaction by the truncation in \vk{}-space, which is a side-effect of the truncation.
Finally, we found converged spectra for the 2\,$\times$\,2\,$\times$\,80 mesh using \ac{RIM} or Coulomb truncation.
Within our spectral resolution of 1\,meV, peaks $E_{nn}$ are at the same places in the spectra for the 2\,$\times$\,2\,$\times$\,100 mesh.

\subsection{\label{sec:bse_cell_size}Cell size convergence}

\begin{figure}[h]
\includegraphics[width=0.99\columnwidth]{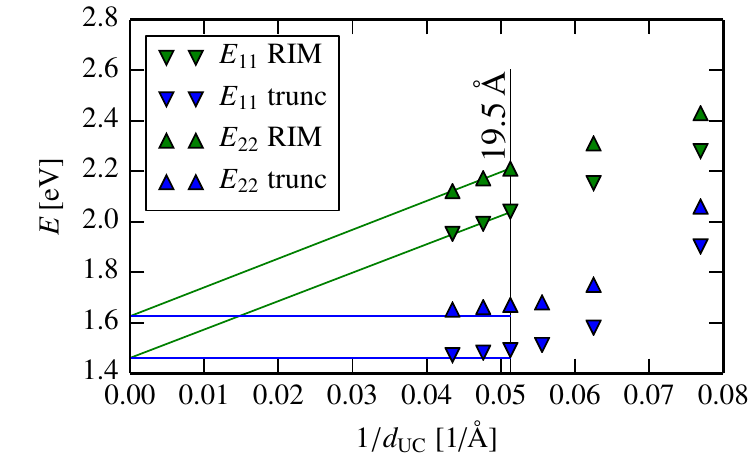}
\caption{\label{fig:bse_cell_conv}(Color online.) Convergence of the first optical transition ($E_{11}$). 
The convergence with Coulomb truncation is much faster than without Coulomb truncation, using \ac{RIM}. The result for the infinite cell is approximately the same.
}
\end{figure}

As the cell size convergence requires additional attention due to the strong Coulomb interaction, the \ac{BSE} calculations also demand for this test.
The cell size test is performed with Yambo using a 2\,$\times$\,2\,$\times$\,40 \vk{}-point sampling, which is sufficiently precise for this purpose.

Figure \ref{fig:bse_cell_conv} depicts the peak position of the first and the second excitonic peak as a function of the inverse cell size.
BSE results without \ac{RIM} or Coulomb truncation are not shown, since the transitions do not converge with increasing number of \vk{}-points.
In the case of \ac{RIM}, the optical transitions converge approximately towards the same value than in the Coulomb truncated case for an extrapolated, infinite cell, however, the extrapolation error can be large.
As expected, Coulomb truncation leads to a significantly improved convergence.

\section{\label{sec:bseassign}Assignment of optical transitions and electronic states}

\begin{figure}[h]
\includegraphics[width=0.99\columnwidth]{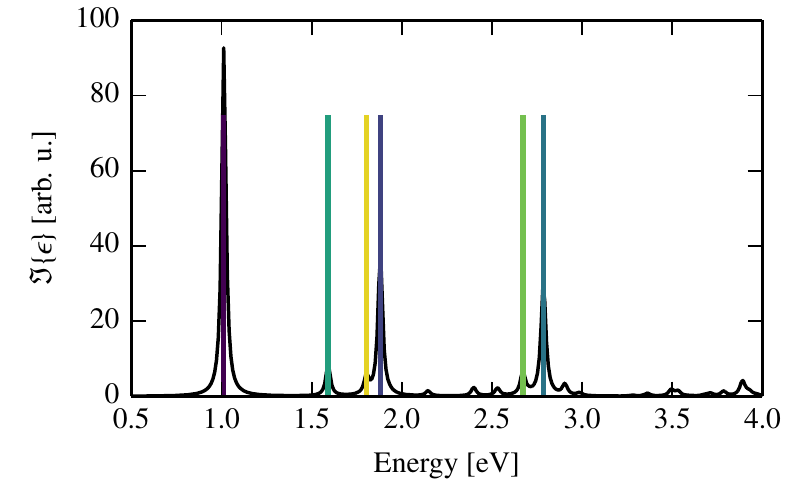}
\caption{\label{fig:exc_ana_spec}(Color online.) The optical spectrum of the (8,0)-CNT (black line) and the six most significant eigenvalues (colored).}
\end{figure}

\begin{figure}[t]
\includegraphics[width=0.99\columnwidth]{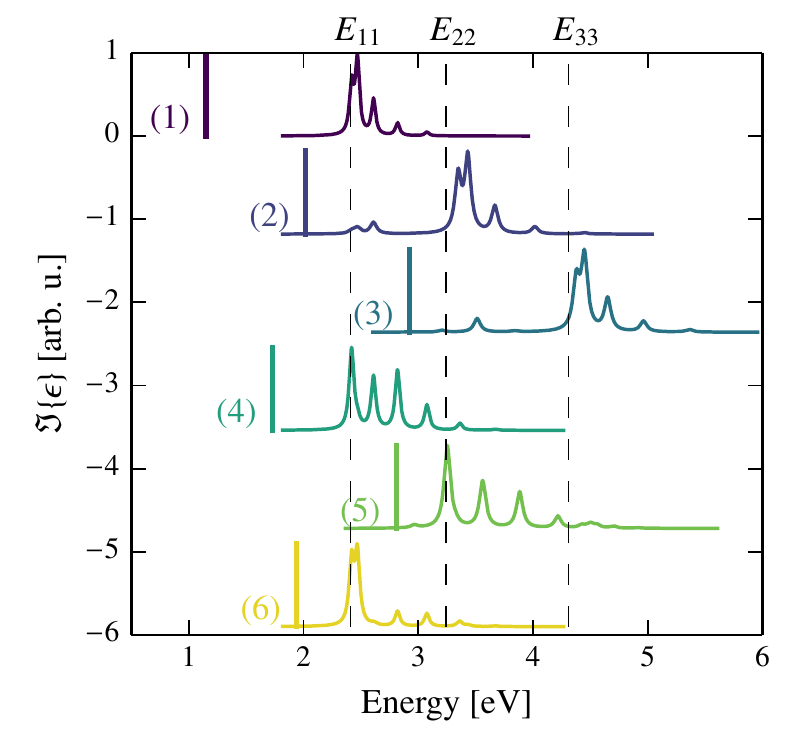}
\caption{\label{fig:exc_ana_trans}(Color online.) The most significant eigenvalues of the exciton Hamiltonian for the (8,0)-CNT (colored bars), the non-interacting Kohn-Sham contributions to each eigenstate (colored lines) and the first, second, and third non-interacting electronic transition (black, dashed lines). Spectra are shifted in $y$ direction for clarity.}
\end{figure}

For computing optical-absorption spectra and dielectric functions, Yambo, by default, implements the efficient Haydock-solver, that does not require explicit knowledge of excitonic eigenvalues and eigenstates \cite{Marini_2009}.
For the assignment of particular optical peaks to single-particle electronic transitions, however, the corresponding eigenstate of the exciton Hamiltonian needs to be known.
This is usually done using iterative diagonalization, up to the excitonic state of interest, since a full diagonalization for a fully converged parameter set would lead to an enormous computational effort.
In order to reduce the numerical effort to a reasonable amount, a reduced  2\,$\times$\,2\,$\times$\,40 ${\vk}$-point set is used, the number of bands for the screening function is reduced to 92, and local-field effects are only treated up to 40 eV.
As a result, the peaks in the optical spectrum in Fig.\ \ref{fig:exc_ana_spec} are located at slightly different positions than in the main text, Fig. \ref{fig:result_bse_spectrum}.
However, the assignment of optical and electronic states is nevertheless transferable.
The colored lines in Fig.\ \ref{fig:exc_ana_spec} correspond to the six most dominant excitonic transitions and the corresponding colors in Fig.\ \ref{fig:exc_ana_trans} refer to the same peaks.

The optical transitions labeled (1), (2), and (3) in Fig.\ \ref{fig:exc_ana_trans} belong to the electronic transitions $E_{11}$, $E_{22}$, and $E_{33}$.
The optical transitions (4) and (6) (dark green, yellow) correspond to the same electronic transition as (1), but with a different distribution in energy and, thus, different extent in $\vk$-space due to the monotonic band structure $E(k)$.
Therefore, these two states are $n$=2 and $n$=3 excitons of the same electronic transition $E_{11}$.
The optical transitions (2) (blue) and (5) (light green) belong to the same electronic transition $E_{22}$. 
The transition (3) (light blue) belongs to the electronic transition $E_{33}$. 

\subsection{Comparison of the BSE results in Yambo and VASP}
\label{app:yambo_vasp}

\begin{figure}[t]
\includegraphics[width=0.99\columnwidth]{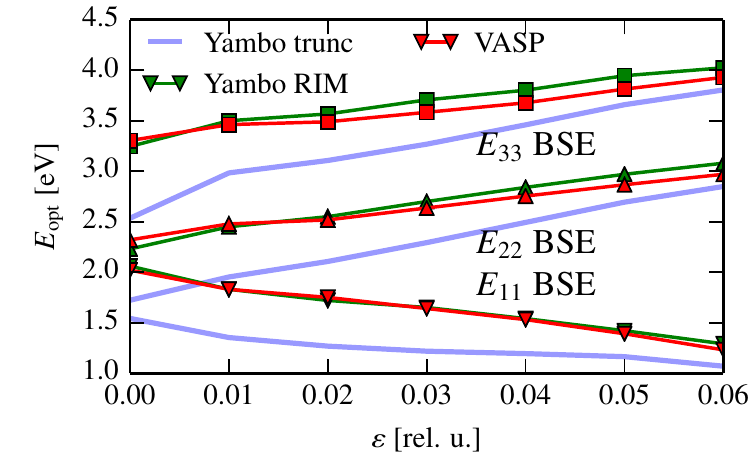}
\caption{\label{fig:bse_yambo_vs_vasp}(Color online.) The first three optical transitions computed using the BSE framework in VASP with homogeneous screening (``VASP'') or Yambo using \ac{RIM} and homogeneous screening (``Yambo RIM'') for the unit-cell size of 19.5\,\AA{}. For comparison, the result with Coulomb truncation is included (``Yambo trunc'').}
\end{figure}

For comparison, we also used the implementation of the \ac{BSE} in VASP described in Refs.\ \onlinecite{Roedl:2008,Fuchs_2008}, which uses a different solver.
The VASP calculations rely on an analytic model to describe the $\vq$-dependence (local fields) of the dielectric function, along with a dielectric constant from the \ac{IP} approximation \cite{Cappellini_1993}, as depicted in Fig.\ \ref{fig:eps_q_approx}.
These results are compared to Yambo where $\epsilon(\vq)$ is calculated for all $\vq$-points using \ac{RPA} (without Coulomb truncation and using the \ac{RIM} for singularity correction).
Figure \ref{fig:bse_yambo_vs_vasp} shows this comparison, and also compares to results computed using the Coulomb truncation.
In Yambo, the interpolated QP energies from $GW$ calculations with Coulomb truncation are used to solve the \ac{BSE}.
In VASP, only the QP gap from the Coulomb truncated $GW$ calculations is used for solving the \ac{BSE}. 
The resulting spectrum is then stretched, according to the band stretching parameters from the main text.

The results of both codes agree almost perfectly for the untruncated case, despite the analytic treatment of the $q$-dependence in the model dielectric function, which behaves similarly for small $q_z$, as long as the same dielectric constant is used (see Fig.\ \ref{fig:eps_q_approx}).
The reason is the large extent of the exciton in real space (at least eight unit cells), concluded from a plot of the exciton wave functions of the first three excitons $E_{11}, E_{22}$, and $E_{33}$, and, thus, its localization in $k$ space.
From this we conclude that the $q_z$-dependence of $\epsilon(q_z)$ for \emph{small} $q_z$ has a much larger impact on the exciton binding energy than the behavior for large $q_z$.

\section{\label{sec:bse_scale}Different screening functions and the scaling parameter $\alpha$}

\begin{figure}[!t]
\includegraphics[width=0.99\columnwidth]{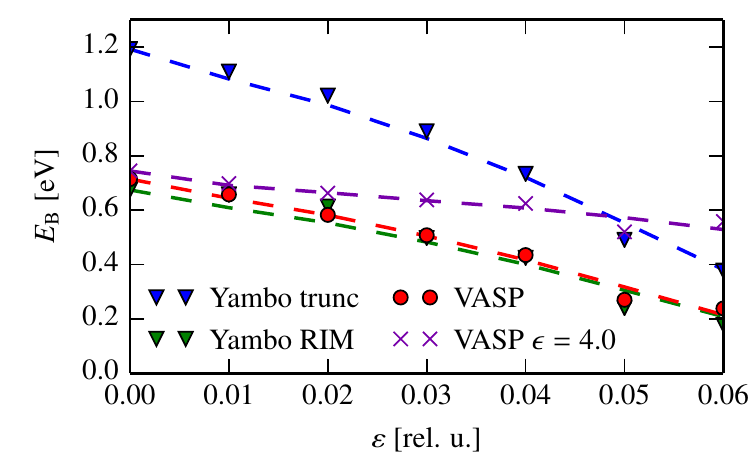}
\caption{\label{fig:bse_scaling_strain_appendix}(Color online.) Scaling of the strain-dependent exciton-binding energy $E\sub{B}$ of the first transition ($E_{11}$) with (``Yambo trunc'') and without Coulomb truncation (``Yambo RIM'', ``VASP'', and ``VASP $\epsilon$=4.0''). 
Symbols represent the result of \ac{BSE} calculations and dashed lines represent the scaling relation, Eq.\ \eqref{eq:bse_scaling} with $\alpha$ obtained from the fit to \ac{BSE} data.}
\end{figure}

In the main text (see Fig.\ \ref{fig:bse_scaling_strain}), the value of $\alpha$ from Eq.\ \eqref{eq:bse_scaling} is discussed by depicting the exciton-binding energy. 
Since the data from Yambo (``Yambo RIM'') and VASP (``VASP'') are almost the same, the VASP BSE results are not discussed explicitly in the main text, but are shown, here (see Fig.\ \ref{fig:bse_scaling_strain_appendix}).
Again, both codes show a good agreement for the exciton-binding energy.
This also reflects in the scaling behavior, which is identical and the value of $\alpha$ of about 1.3 is the same for both cases.

Since in the VASP code, the dielectric constant used in the model dielectric function can be specified, this allows us to independently investigate the scaling with respect to the effective mass.
We do this here by fixing the dielectric constant at a value of 4.0, as depicted in Fig.\ \ref{fig:bse_scaling_strain_appendix}.
The violet data shows that the change of the binding energy with respect to the effective mass is much smaller than with respect to the dielectric constant.
This also becomes clear from the scaling relation (main text, Eq.\ \eqref{eq:bse_scaling}) and we find by fitting that the same value for $\alpha = 1.29\pm0.03$ is obtained.

\section{Electronic transitions of DFT(HSE06) and optical transitions of the BSE}

Barone \textit{et al.}\ state that it is possible to approximately predict the optical transitions of \acp{CNT} by calculating their electronic transitions using \ac{DFT} employing the unmodified HSE06 functional, labeled DFT(HSE06) \cite{Barone_2011}.
This approach works for materials for which the exciton binding energy and \ac{QP} corrections of excited states almost counter-correct each other.
However, it does not easily work for strain-dependent optical transitions in \acp{CNT}, as we explain in the following.

\begin{figure}[t!]
\includegraphics[width=0.99\columnwidth]{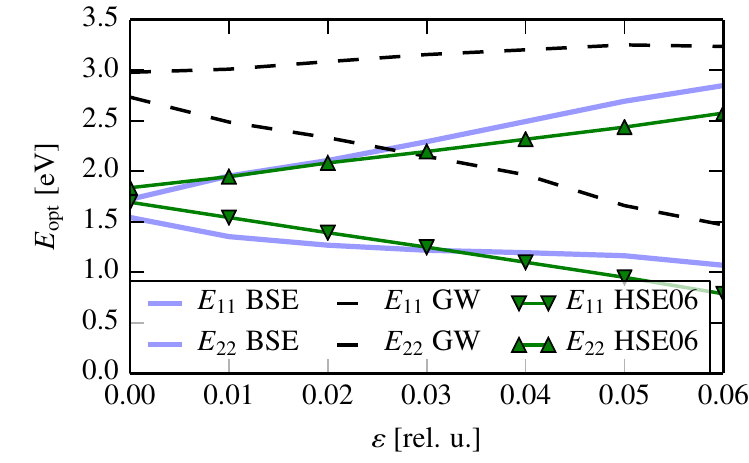}
\caption{\label{fig:bse_vs_hse} (Color online.) The comparison of the first ('$E_{11}$') and second ('$E_{22}$') optically active band transition using HSE06 ('HSE06'), the $GW$ approximation ('GW') and the first and second optical transition in the BSE framework ('BSE').}
\end{figure}

Fig. \ref{fig:bse_vs_hse} shows the first two electronic transitions of the (8,0)-CNT calculated by three methods:
Fully solving the \ac{BSE} on top of the \acs{GW} approximation ('BSE'), only solving the \ac{GW} approximation ('GW'), and using DFT(HSE06) ('HSE06' in the figure).
It appears that the optically active electronic transitions of the (8,0)-CNT calculated by DFT(HSE06) are closer to the optical transitions from the \ac{BSE} solution than to the electronic $GW$ transitions.
However, the strain dependent shifts of the electronic transitions, calculated by DFT(HSE06) and $GW$, are very similar --- whereas the optical transitions from the solution of the \ac{BSE} shift differently with strain. 
The reason for this result are the screening effects: Both, the difference between the \ac{QP} correction and the exciton binding energy depends on strain, which cannot be captured by \ac{DFT}(HSE06) calculations alone.

%% file: preamble.tex
\usepackage{dcolumn} 		
\usepackage{bm}
\usepackage[utf8]{inputenc} 
\usepackage{color}          
\usepackage{hyperref}       
\usepackage{graphicx}       
\usepackage{url}            
\usepackage{acro}    		
\usepackage{enumitem} 		
\usepackage[normalem]{ulem} 
\usepackage[mathlines]{lineno}
\usepackage{xr}

\renewcommand{\vec}[1]{\ensuremath{\mathbf{#1}}}

\newcommand{\vG}{\ensuremath{\vec{G}}}
\newcommand{\vk}{\ensuremath{\vec{k}}}
\newcommand{\vq}{\ensuremath{\vec{q}}}

\newcommand{\sub}[1]{\ensuremath{_\mathrm{#1}}}
\newcommand{\up}[1]{\ensuremath{^\mathrm{#1}}} 

\newcommand{\sbr}[1]{\ensuremath{\left[#1\right]}} 

\definecolor{dblue}{rgb}{0, 0, 0.7}
\definecolor{dgreen}{rgb}{0, 0.5, 0}
\hypersetup{
pdftitle = {Strain and quasiparticle excitations: Optical properties of carbon nanotubes from first principles}, 
colorlinks = true,
linkcolor = dblue,
urlcolor = blue,
citecolor = dgreen,
}

\definecolor{orange}{rgb}{0.7, 0.35, 0.0}

%% file: acronyms.tex
\DeclareAcronym{1D}{
  short = 1D,
  long = one dimensional,
  class = abbrev
}

\DeclareAcronym{2D}{
  short = 2D,
  long = two dimensional,
  class = abbrev
}

\DeclareAcronym{3D}{
  short = 3D,
  long = three dimensional,
  class = abbrev
}

\DeclareAcronym{AI}{
  short = AI,
  long = \textit{ab initio},
  class = hidden
}

\DeclareAcronym{BSE}{
  short = BSE,
  short-plural = s,
  long = Bethe-Salpeter equation,
  long-plural = s,
  class = abbrev
}

\DeclareAcronym{BZ}{
  short = BZ,
  long = Brillouin zone,
  class = abbrev
}

\DeclareAcronym{CB}{
  short = CB,
  long  = conduction band,
  class = abbrev
}

\DeclareAcronym{CNT}{
  short = CNT,
  long = carbon nanotube,
  class = abbrev
}

\DeclareAcronym{DF}{
  short = DF,
  long  = dielectric function,
  class = abbrev
}

\DeclareAcronym{DFT}{
  short = DFT,
  long  = density functional theory,
  class = abbrev
}

\DeclareAcronym{DFTB}{
  short = DFTB,
  long  = density functional based tight binding,
  class = abbrev
}

\DeclareAcronym{ES}{
  short = ES,
  long  = electronic structure,
  class = abbrev
}

\DeclareAcronym{etal}{
  short = \emph{et al.},
  long = \emph{et al.},
  class = hidden
}

\DeclareAcronym{EXX}{
  short = EXX,
  long  = exact exchange,
  class = abbrev
}

\DeclareAcronym{GGA}{
  short = GGA,
  long  = generalized gradient approximation,
  class = abbrev
}

\DeclareAcronym{GW}{
  short = $GW$,
  long = $GW$ approximation,
  class = abbrev
}

\DeclareAcronym{G0W0}{
  short = $G_0W_0$,
  long = $G0W0$ approximation,
  class = abbrev
}

\DeclareAcronym{HF}{
  short = HF,
  long  = Hartree-Fock,
  class = abbrev
}

\DeclareAcronym{HK}{
  short = HK,
  long  = Hohenberg-Kohn,
  class = abbrev
}

\DeclareAcronym{HOMO}{
  short = HOMO,
  long  = highest occupied molecule orbital,
  class = abbrev
}

\DeclareAcronym{IP}{
  short = IP,
  long  = independent particle,
  class = abbrev
}

\DeclareAcronym{KS}{
  short = KS,
  long  = Kohn-Sham,
  class = abbrev
}

\DeclareAcronym{LDA}{
  short = LDA,
  long  = local-density approximation,
  class = abbrev
}

\DeclareAcronym{lhs}{
  short = lhs,
  long  = left hand side,
  class = abbrev
}

\DeclareAcronym{LUMO}{
  short = LUMO,
  long  = lowest unoccupied molecule orbital,
  class = abbrev
}

\DeclareAcronym{mCNT}{
  short = mCNT,
  short-plural = s,
  long = metallic \ac{CNT},
  long-plural = s,
  class = abbrev
}

\DeclareAcronym{MB}{
  short = MB,
  long  = many-body,
  class = abbrev
}

\DeclareAcronym{MD}{
  short = MD,
  long  = molecular dynamics,
  class = abbrev
}

\DeclareAcronym{MEMS}{
  short = MEMS,
  long  = micro electro-mechanical systems,
  class = abbrev
}

\DeclareAcronym{MOEMS}{
  short = MOEMS,
  long  = micro-opto-electro-mechanical systems,
  class = abbrev
}

\DeclareAcronym{MM}{
  short = MM,
  long  = molecular mechanics,
  class = abbrev
}

\DeclareAcronym{MP}{
  short = MP,
  long  = Monkhorst-Pack,
  class = abbrev
}

\DeclareAcronym{mwCNT}{
  short = mwCNT,
  long = multi-walled \ac{CNT},
  long-plural = ,
  class = abbrev
}

\DeclareAcronym{PDE}{
  short = PDE,
  long  = partial differential equation,
  class = abbrev
}

\DeclareAcronym{PL}{
  short = PL,
  long  = photoluminescence,
  class = abbrev
}

\DeclareAcronym{PPA}{
  short = PPA,
  long  = plasmon-pole approximation,
  class = abbrev
}

\DeclareAcronym{PW}{
  short = PW,
  long  = plain wave,
  class = abbrev
}

\DeclareAcronym{QMC}{
  short = QMC,
  long  = quantum Monte Carlo,
  class = abbrev
}

\DeclareAcronym{QP}{
  short = QP,
  long  = quasiparticle,
  class = abbrev
}

\DeclareAcronym{RIM}{
  short = RIM,
  long  = random-integration method,
  class = abbrev
}

\DeclareAcronym{RPA}{
  short = RPA,
  long  = random phase approximation,
  class = abbrev
}

\DeclareAcronym{SE}{
  short = SE,
  long  = Schrödinger equation,
  class = abbrev
}

\DeclareAcronym{scCNT}{
  short = scCNT,
  long = semi conducting \ac{CNT},
  class = abbrev
}

\DeclareAcronym{SIC}{
  short = SIC,
  long  = self interaction correction,
  class = abbrev
}

\DeclareAcronym{scGW}{
  short = scGW,
  long  = self consistent \ac{GW},
  class = abbrev
}

\DeclareAcronym{swCNT}{
  short = swCNT,
  long = single-walled \ac{CNT},
  long-plural = ,
  class = abbrev
}

\DeclareAcronym{TB}{
  short = TB,
  long  = tight binding,
  class = abbrev
}

\DeclareAcronym{VB}{
  short = VB,
  long  = valence band,
  class = abbrev
}

\DeclareAcronym{VdW}{
  short = VdW,
  long  = Van-der-Waals,
  class = abbrev
}

\DeclareAcronym{wrt}{
  short = wrt.\ ,
  long  = with respect to,
  class = abbrev
}

\DeclareAcronym{XC}{
  short = XC,
  long  = exchange and correlation,
  class = abbrev
}